\documentclass[12pt]{article}
\usepackage{amsmath,amssymb,array}
\numberwithin{equation}{section}
\def \be{\begin{equation}}
\def \ee{\end{equation}}
\def \bea{\begin{eqnarray}}
\def \eea{\end{eqnarray}}
\def \p{\partial}

\def \a{\alpha}

\def \G{\Gamma}

\begin{document}

\begin{titlepage}
\renewcommand{\thefootnote}{\fnsymbol{footnote}}

\begin{center}

\begin{flushright}
hep-th/0509113
\end{flushright}
\vspace{1.5cm}

\textbf{\Large{Exactly Solvable Model of Superstring \\[0.5cm]
in Plane-wave Background with Linear Null Dilaton}}\\[1.5cm]

\large{Bin Chen\footnote{Email: bchen01@pku.edu.cn}\hspace{0.5cm}
Ya-Li He\footnote{Email: ylhe@pku.edu.cn}}\\[0.5cm]
\emph{Department of Physics, Peking University, \\
      Beijing 100871, P.R. China} \\[1cm]
\large{Peng Zhang\footnote{Email: pzhang@itp.ac.cn}}\\[0.5cm]
\emph{Institute of Theoretical Physics, Chinese Academy of Science, \\
      P.O. Box 2735, Beijing 100080, P.R. China}

\end{center}
\vspace{1cm}

\centerline{\textbf{Abstract}}\vspace{0.5cm}

In this paper, we study an exactly solvable model of IIB
superstring in a time-dependent plane-wave backgound with a
constant self-dual Ramond-Ramond 5-form field strength and a
linear dilaton in the light-like direction. This background keeps
sixteen supersymmetries. In the light-cone gauge, the action is
described by the two-dimensional free bosons and fermions with
time-dependent masses. The model could be canonically quantized
and its Hamiltonian is time-dependent with vanishing zero-point
energy. The spectrum of the excitations is symmetric between the
bosonic and fermionic sector. The string mode creation turns out
to be very small.

\end{titlepage}
\setcounter{footnote}{0}


\section{Introduction}

The study of string theory in the time-dependent backgrounds has
attracted much attention in the past decade. Similar to quantum
field theory in the curved spacetime, string theory in the
time-dependent backgrounds shares the same unclear subtle issues
such as the choice of vacuum, string creation et al. Not
mentioning a second-quantized version, even a perturbative
description of string theory in a time-dependent background is in
general out of control. Nevertheless, the issue is essential for
us to understand the fundamental problems in quantum gravity. One
central problem in quantum gravity is the resolution of the cosmic
singularity. People have been expecting that the big-bang
singularity could be resolved in string theory, namely the
cosmological singularity should not exist in string theory. Unlike
the usual orbifold or conifold singularity, the big-bang
singularity is space-like and its resolution requires the
knowledge of string theory in a time-dependent background.

In the past few years, there appeared a few remarkable models to
address the issue. One class of such models is the time-dependent
orbifolds \cite{GLS,Cornalba}. These models are solvable and keep
part of supersymmetries. Unfortunately it turns out that the null
orbifold background is unstable due to the large blueshift effect
\cite{HP, Lawrence} and the perturbative string theory in many
time-dependent orbifolds breaks down \cite{GLS1, Kutasov}. The
$\alpha^\prime$ correction and $g_s$ correction play important
roles near the singularity.  Nevertheless, the recent development
shows that the twisted states condensation may induce an RG-flow
from null orbifold to the usual spatial orbifold \cite{Berkooz}.
Another class of solvable time-dependent models is the rolling
tachyon, which shows the features such as tachyon condensation,
unusual open/close string duality \cite{Sen}. Very recently,
several new ideas have been proposed. In \cite{Silverstein}, the
authors argued that there is a closed string tachyon condensation
phase replacing the cosmological singularity. In this case, the
perturbative string amplitudes are self-consistently truncated to
the small coupling region such that they are finite, indicating a
resolution of  the singularity.

Another interesting idea is so called matrix big bang. In
\cite{Verlinde}, it has been proposed that in a linear null
dilaton background it is the matrix degrees of freedom rather than
the point particles or the perturbative strings that describe the
physics near the big-bang singularity correctly. Some related
discussions can be found in \cite{Miao, Bin, Das, she}, see also
\cite{o1,o2}. The linear null dilaton background is different from
the usual linear (spatial) dilaton. One obvious difference is that
the latter one always leads to a noncritical string theory, while
the former one does not modify the central charge or the dimension
of the original theory. The common feature of two backgrounds is
that due to the linearity of the dilaton, the string coupling
becomes very strong in some region where the perturbative string
theory is not valid. In the linear spatial dilaton case, one can
add a tachyon background (Liouville wall) to suppress the
contribution coming from the strong coupling region in the path
integral such that the perturbative theory is
well-defined\cite{Liouville}. Due to the existence of the
Liouville wall, the incoming wave is reflected and becomes
outgoing. Among the theories with Liouville field, the 2D
noncritical string theory with a time direction and a Liouville
direction is of particular interest (See nice reviews
\cite{2Dstring}). It has a dual description by matrix model, which
is exactly solvable. Actually, matrix model describes the dynamics
of the D0-particles in 2D string theory\cite{Matrix}. This
impressive relation between 2D string theory and matrix model is
an illuminating manifestation of the open/closed duality. The
story is a little different in the linear null dilaton case. It
seems that one does not have a tachyon field to make the
perturbative string theory well-defined. Instead, with the similar
spirit leading to BFSS matrix model\cite{BFSS}, a matrix model has
been proposed to describe the physics in the strong coupling
region\cite{Verlinde}. Similarly, the matrix degrees of freedom
are identified with the open string degrees of freedoms between
the partons.

Inspired by the recent interest in the linear null  dilaton
background, we will study an exactly solvable string model in a
plane-wave background in the presence of self-dual
Ramond-Ramond(RR) 5-form field strength and a linear dilaton in
the light-cone direction. The study of string theory in a
plane-wave background has a long history. It is remarkable that
\cite{90Horow} the plane-fronted waves are not only the solutions
of the effective supergravity but even the exact solutions of the
string theory. Furthermore in \cite{Metsaev01, MT02} the IIB
superstring in the plane-wave background with a constant self-dual
Ramond-Ramond field strength and a constant dilaton is quantized
in the lightcone gauge. This leads to the famous PP-wave/SYM
correspondence\cite{BMN}. Later on, a time-dependent plane-wave
string model was studied in \cite{02Tsey}.

In this paper we consider the type IIB Green-Schwarz superstring in
the following time-dependent background with Ramond-Ramond flux:
\begin{eqnarray}
&&ds^2 = -2dx^+dx^--\lambda(x^+)\,x_I^2\,dx^+dx^+ +dx^Idx^I\,,\nonumber \\
&&\phi=\phi(x^+)\,,\quad\quad(F_5)_{+1234}=(F_5)_{+5678}=2f.\label{BG}
\end{eqnarray}
The world-sheet conformal invariance requires
\begin{eqnarray}\label{Conformal}
R_{\mu\nu}=-2D_{\mu}D_{\nu}\phi+\frac{1}{24}e^{2\phi}(F^2_{5})_{\mu\nu}.
\end{eqnarray}
The only nonzero component of the Ricci curvature tensor
$R_{\mu\nu}$ with respect to the metric in (\ref{BG}) is
$R_{++}=8\lambda(x^+)$. Inserting (\ref{BG}) into the conformal
invariance condition (\ref{Conformal}), we get the relation
\begin{eqnarray}
\lambda=-\frac{1}{4}\phi''+f^2e^{2\phi}\,.
\end{eqnarray}
From now on we restrict the dilaton to be linear in the light-cone
time coordinate $x^+$, i.e. $\phi=-cx^+$ with $c$ being a constant,
then we have
\begin{eqnarray}\label{lam}
\lambda=f^2e^{-2cx^+}.
\end{eqnarray}
In general, $f$ could be an arbitrary function of $x^+$ and there
is a large class of the models. These models could be studied in
the lightcone gauge. For a generic $f$, it is difficult to solve
the model analytically. However,  there are two special cases:
\begin{itemize}
\item One is that $\lambda$ is a constant. This happens when
$f=f_0e^{-\phi}$ with $f_0$ being constant so that
 \be \label{PP}
 \lambda=f^2_0\,,
 \ee
and the metric in the string frame reduces to the form of the
maximally supersymmetric plane wave \cite{Blau}. Actually string
theory in this background is very similar to the one in the
maximally supersymmetric case. The bosonic and fermionic action,
the quantization and the spectrum are all the same. However, the
vertex operators and the perturbative amplitudes are different due
to the existence of the linear dilaton. It could be expected that
the perturbative string theory is not well-defined in the strong
coupling region and an alternative description, say the matrix
degrees of freedom, is needed.

\item The other special case is when the self-dual RR field
strength $f=f_0$ is constant. This is the case we will pay most of
our attention to in this paper. In this case, the perturbative
string theory is still exactly solvable. We manage to quantize
both the bosonic and fermionic sectors of string theory in the
light-cone gauge. It turns out that the Hamiltonian is
time-dependent and the zero-point energy cancels between the
bosonic and fermionic sectors. And the spectrum of the bosonic and
fermionic excitations is symmetric. Similarly the string coupling
is very strong near the big bang singularity.
\end{itemize}

Note that without loss of generality the coefficient $c$ in the
dilaton can be set to any value  by rescaling the coordinates
$x^+,x^-$ and the RR field strength $f$ properly. In the following
sections when we solve this model in the light-cone gauge, we will
set $c=\frac{1}{\alpha'p^+}$ for  simplicity. In this case, $c$ is
positive and indicates that the strong string coupling region is
near $x^+ \rightarrow -\infty$.

The paper is organized as follows. In section 2, we investigate
some properties of the background: its symmetry algebra, the
geodesic incompleteness, supersymmetries and its description in
Rosen coordinates. In section 3, we study the quantization of the
bosonic sector. In section 4, we turn to the fermionic sector,
which needs more techniques. In section 5, we discuss the quantum
string mode creation. We end the paper with the conclusion and
some discussions.

\section{Some Properties of the Background}

Let us start from the metric in (\ref{BG}) and discuss some
properties of the background.

\subsection{Symmetry Algebra of the background}

We are interested in the symmetry algebra of the background, which
is encoded in the complete set of Killing vectors preserving the
background (\ref{BG}).  It is manifest that the background is
invariant under translation in the $x^{-}$ direction, and the
corresponding Killing vector is
\begin{eqnarray}
T\equiv\,\frac{\partial}{\partial x^{-}}\,,
\end{eqnarray}
which generates a $\mathbb{R}$-subgroup of the total isometry group.
But unlike the maximally symmetric case in \cite{Blau}, the
translation invariance in the $x^+$ direction is broken by the
nontrivially $x^+$-dependent dilaton. Although the metric does not
possess $SO(8)$ symmetry due to the presence of RR five-form flux,
the background is invariant under two $SO(4)$'s, which act on $x^i$
and $x^a$ directions separately. We denote the corresponding
generators by $J_{ij}$ and $J_{ab}$ with $i,j$ running from 1 to 4
and $a,b$ from 5 to 8,
\begin{eqnarray}
J_{ij}\equiv x_i\partial_j-x_j\partial_i\,\,\,\,\,\,\,,
\,\,\,\,\,\,\,J_{ab}\equiv x_a\partial_b-x_b\partial_a\,\,.
\end{eqnarray}
If we change these two $SO(4)$'s, the solution is still invariant.
So we get a $\mathbb{Z}_2$ symmetry,
\begin{eqnarray}
\{x^i\} \longleftrightarrow \{x^a\}.
\end{eqnarray}

It is evident that the translations along $x^I=\{x^i,\,x^a\}$ do not
leave the solution invariant. However, if we shift $x^-$
appropriately at the same time, we find new symmetric translations
with the generators
\begin{eqnarray}\label{L_I}
L_I=a\partial_I+(\partial_+a)\,x_I\partial_-\,,
\end{eqnarray}
with $a(x^+)$\,satisfying
\begin{eqnarray}
\partial_+^2a+\lambda \,a=0\,.\label{AA}
\end{eqnarray}
Since (\ref{AA}) is a 2-order equation, we get two sets of
generators denoted by $L_I$, $\tilde{L}_I$.
\begin{enumerate}
\item In the case $\lambda=f^2_0$, we have
\begin{eqnarray}
L_I&=&\cos(f_0x^+)\p_I-f_0\sin(f_0x^+)x_I\partial_{-}\,,\nonumber\\
\tilde{L}_I&=&\sin(f_0x^+)\p_I+f_0\cos(f_0x^+)x_{I}\partial_{-}\,.
\end{eqnarray}

\item In the case $\lambda=f^2_0\,e^{-2cx^+}$, we get
\begin{eqnarray}
L_I&=&J_0\left(\frac{f_0}{c}\,e^{-cx^+}\right)\partial_I+
                  f_0\,e^{-cx^+}J_1\left(\frac{f_0}{c}\,e^{-cx^+}\right)x_{I}\partial_{-}\,,\nonumber\\
\tilde{L}_I&=&Y_0\left(\frac{f_0}{c}\,e^{-cx^+}\right)\partial_I+
                  f_0\,e^{-cx^+}Y_1\left(\frac{f_0}{c}\,e^{-cx^+}\right)x_{I}\partial_{-}\,.
\end{eqnarray}

\end{enumerate}

It is straightforward to write down the non-vanishing commutators as
follows :
\begin{eqnarray}
&&[L_{I},\,\,\tilde{L}_{J}]\,=\gamma T\delta_{IJ}\,,\nonumber\\
&&[J_{IJ},\,L_{K}]=L_{I}\delta_{JK}-L_{J}\delta_{IK}\,,\nonumber\\
&&[J_{IJ},\,\tilde{L}_K]=\tilde{L}_I\delta_{JK}-\tilde{L}_J\delta_{IK}\,,\nonumber\\
&&[J_{IJ},J_{KL}]=J_{IL}\delta_{JK}+J_{JK}\delta_{IL}-J_{IK}\delta_{JL}-J_{JK}\delta_{IL}\,,
\end{eqnarray}
where \be
 \gamma=\left\{\begin{array}{ll}
 f_0,& \mbox{when $\lambda=f_0^2$,} \\[0.1cm]
 -\frac{2c}{\pi}, & \mbox{when $\lambda=f_0^2\,e^{-2cx^+}$.}
 \end{array}\right.
 \ee
Here we have used the identity
$J_0(z)Y_1(z)-J_1(z)Y_0(z)=-\frac{2}{\pi z}$ to get the first
commutator in the second case and for simplicity we have defined
$J_{IJ}=\{J_{ij},J_{ab}\}$. From the commutators above, we can see
$L_I$, $\tilde{L}_I$ and $T$ form the Heisenberg-type algebra $h(8)$
with $\hbar$ being replaced by $\gamma$. Therefore in both cases the
continuous symmetry algebra is $[so(4)\oplus so(4)]\oplus_sh(8)$.
Here $\oplus_s$ means semi-direct sum. So the background (\ref{BG})
admits a symmetry algebra of dimension twenty-nine.

Another way to check the symmetry algebra is to work in the Einstein
frame and figure out the Killing vectors which keep the RR
background invariant. Following the discussion in \cite{Bin}, it is
straightforward to find that the Killing vector corresponding to the
translational invariance along $x^+$ does not exist, which indicates
the absence of the supernumerary supersymmetries. In fact, the
Killing symmetries discussed above are the complete set preserving
the background. It is not hard to check that our background is
inhomogeneous.

With the above Killing vectors, there are a few physical
implications. Firstly one may define the conserved quantities with
the Killing vectors. For instance, the conserved quantity
corresponding to the translational invariance Killing vector $T$ is
the lightcone momentum. On the other hand, the absence of the
Killing vector $\p_+$ indicates that the lightcone Hamiltonian is
not conserved. Actually we will show that the Hamiltonian in our
background is time-dependent. Another implication is that due to the
existence of the null Killing vector $T$, there is no particle or
string creation in our background \cite{Gibbons}.

\subsection{Geodesics and Tidal force}
In this subsection we discuss some geometric features of our background
according to the general method of \cite{Marolf}.
Using the metric in our background (\ref{BG}), we can obtain the
non-vanishing components of the Christoffel connection
\begin{eqnarray}\label{Chr}
\Gamma^-_{++}=-c\lambda
x^2_I\quad\,,\quad\Gamma^-_{+I}=\Gamma^I_{++}=\lambda x_I\quad\,.
\end{eqnarray}
It is easy to get the nonzero components of the Riemann tensor
\begin{eqnarray}
R_{I+I+}=\lambda\,\,.
\end{eqnarray}
In our background the geodesic equation
\begin{eqnarray}
\frac{d^2x^{\mu}}{d\sigma^2}+\Gamma^{\mu}_{\alpha\beta}\frac{dx^{\alpha}}{d\sigma}\frac{dx^{\beta}}{d\sigma}=0
\end{eqnarray}
can be rewritten as
\begin{eqnarray}\label{GeE}
&&\frac{d^2x^{+}}{d\sigma^2}=0\,\,,\nonumber\\
&&\frac{d^2x^{-}}{d\sigma^2}+ 2\lambda
x_I\frac{dx^{+}}{d\sigma}\frac{dx^{I}}{d\sigma}
                 -c\lambda x^2_I\frac{dx^{+}}{d\sigma}\frac{dx^{+}}{d\sigma}=0\,\,,\nonumber\\
&&\frac{d^2x^{I}}{d\sigma^2}+ \lambda
x_I\frac{dx^{+}}{d\sigma}\frac{dx^{+}}{d\sigma}=0\,\,,
\end{eqnarray}
where $\sigma$ is an affine parameter.

The general solution of the first equation in (\ref{GeE}) is
\begin{eqnarray}
x^+=x^+_0\sigma+x^+_1\,\,,
\end{eqnarray}
where $x^+_0$ and $x^+_1$ are constants. If $x^+_0=0$, we get
\begin{eqnarray}
x^+=x^+_1, \quad\,\quad x^-=x^-_0\sigma+x^-_1, \quad\,\quad
x^I=x^I_0\sigma+x^I_1\,\,,
\end{eqnarray}
with $x_0$ and $x_1$ being constants. These geodesics are as in
flat space. If $x^+_1\neq0$, for simplicity we write $x^+=\sigma$,
which is equivalent to rescale $\sigma$ and shift its origin. It
is easy to check that the curve
\begin{eqnarray}\label{curve1}
x^+=\sigma, \quad\,\quad x^-=0, \quad\, \quad x^I=0\,\,,
\end{eqnarray}
is a geodesic. Using the translation $e^{\xi}$ to the curve
(\ref{curve1}), we can obtain a new family of null geodesics. Here
$\xi$ is the Killing field associated with (\ref{L_I}) and
$\xi^{\mu}=(0,\,x^I\partial_+a,\,\xi^I)$ with $\xi^I=a$ defined in
(\ref{AA}). These new geodesics are
\begin{eqnarray}\label{curve2}
x^+=\sigma, \quad\,\quad x^-=\frac{1}{2}\xi^I\partial_+\xi_I,
\quad\,\quad x^I=\xi^I\,\,.
\end{eqnarray}
The distance separating the geodesics (\ref{curve1}) and
(\ref{curve2}) along any surface $x^+=constant$ is given by
$|\xi|=\sqrt{|\xi_I\xi^I|}$. Since $x^+$ is an affine parameter for
this family of geodesics, it is meaningful to speak of the relative
velocity $\partial_+\xi^I$ and acceleration $\partial^2_+\xi^I$ of
these geodesics. Note that these relative accelerations measure
gravitational tidal forces. If the geodesics are infinitesimally
separated, the tidal force is given by certain components of the
Riemann tensor:
\begin{eqnarray}\label{tidal}
\partial^2_+\xi^I=-R^I_{\alpha\beta\gamma}\xi^\beta t^\alpha
t^\gamma\,\,,
\end{eqnarray}
where $t^\alpha$ is the tangent vector to the geodesic. In our
case, (\ref{tidal}) takes the form
\begin{eqnarray}
\partial^2_+a=-\lambda\,a\,\,,
\end{eqnarray}
which is just the equation (\ref{AA}) and $\lambda$ characterizes
the strength of the tidal force. We can see that as
$x^+\rightarrow-\infty$, the tidal force becomes divergent in the
case with constant RR field strength. And since $\lambda$ is
positive, the tidal force is always attractive.

The above discussions are in the string frame. Due to the existence
of the linear null dilaton, the test particle or string will
interact with the dilaton, which drives the test particle away from
the geodesics above. It is better to work in the Einstein frame.
Things become more subtle then.  In the Einstein frame, the metric
in our background (\ref{BG}) can be written as
\begin{eqnarray}\label{metric.E}
ds^2_E=e^{-\frac{\phi}{2}}ds^2_{str.}\,.
\end{eqnarray}
The non-vanishing components of the corresponding Christoffel
connection are
\begin{eqnarray}\label{Chr1}
&&\Gamma^+_{++}=\frac{c}{2} \,\,,\nonumber\\
&&\Gamma^-_{++}=-\frac{5}{4}c\lambda x^2_I\quad\,,
               \quad\Gamma^-_{+I}=\lambda x_I\quad\,,\quad
               \Gamma^-_{II}=\frac{c}{4}\,\,,\nonumber\\
&&\Gamma^I_{++}=\lambda x_I\quad\,,\quad
               \Gamma^I_{+I}=\frac{c}{4}\,\,.
\end{eqnarray}
and the nonzero components of the Riemann tensor take the form
\begin{eqnarray}
R_{I+I+}=\frac{c^2}{16}+\lambda\,\,.
\end{eqnarray}
It's easy to see that the Einstein metric has a singularity at $x^+
\rightarrow -\infty$ since some of metric components go to zero.
Actually such a singularity occurs at the finite geodesic distance,
which indicates the spacetime is geodesically incomplete. Let us
focus on the lines $x^I=0, x^-=\mbox{constant}$, which are
geodesics, and consider the geodesic equation for $x^+$
\begin{eqnarray}\label{x^+}
\frac{d^2x^{+}}{d\sigma^2}+\frac{c}{2}\left(\frac{dx^{+}}{d\sigma}\right)^2=0\,\,,
\end{eqnarray}
which gives
 \be
 e^{\frac{c}{2}x^+}(\frac{dx^+}{d\sigma})=\mbox{constant}.
 \ee
 Hence the affine parameter is
 \be
 \sigma= e^{\frac{c}{2}x^+}
 \ee
 up to an affine transformation. Therefore the singularity
 $x^+\rightarrow -\infty$ corresponds to $\sigma=0$ and it has
 finite affine distance to all points in the interior. In terms of
 the affine parameter $\sigma$, the metric could be rewritten as
 \be
 ds^2_E=-\frac{4}{c}d\sigma
 dx^--\left(\frac{2f}{c}\right)^2\sigma^{-5}x^2_I d\sigma^2+\sigma dx^2_I
 \ee
 and the nonvanishing components of the corresponding Riemann tensor
 are
 \be
 R_{\sigma I \sigma I} \frac{1}{4\sigma^2}+\left(\frac{2f}{c}\right)^2\frac{1}{\sigma^6},
 \ee
which shows a curvature singularity at $\sigma=0$ and gives a
divergent tidal force felt by an inertial observer.

In short, our background (\ref{BG}) is geodesically incomplete and
hence singular from the standpoint of general relativity. This
background is analogue to those with a cosmological singularity.
Near the singularity, the test particle experiences divergent
gravitational tidal force, while for the string the string
interaction becomes extremely strong and the string feels divergent
tidal force. One needs a nonperturbative description of the string
theory near the cosmological singularity.

\subsection{Supersymmetries of the background}

Now let's check  that half of the maximally space-time
supersymmetry has been broken in our background. To this end, we
need to know how many independent Killing spinors the background
(\ref{BG}) supports. The gravitino and dilatino variations should
vanish for independent Killing spinors, i.e.
\begin{eqnarray}
\delta_\epsilon\lambda^A\equiv
(\mathcal{\tilde{D}})^A_B\,\epsilon^B=0\,,\quad
\delta_\epsilon\psi^A_\mu\equiv
(\mathcal{\hat{D}}_\mu)^A_B\,\epsilon^B=0\,,
\end{eqnarray}
where $\mu=+,-,1,...,8$ and $A=1,2$ . The dilatinos $\lambda^A$,
gravitinos $\psi^A_\mu$, and Killing spinors $\epsilon^A$ are all
ten dimensional Weyl-Majorana fermions of positive chirality, so
they carry a 10d spinor index, $\alpha=1,2,\ldots,32$,
additionally, e.g. $\epsilon^A_{\,\alpha}$ etc. (for our notations
and conventions about spinors see Appendix A). In the string
frame, the generalized covariant derivative $\tilde{\mathcal{D}}$
and $\hat{\mathcal{D}}_\mu$ for our background (\ref{BG}) are
\cite{Cvetic1, Cvetic2}
\begin{eqnarray}
(\,\mathcal{\tilde{D}})^A_B
        &=&\frac{1}{2}\,\delta^A_B\,\Gamma^\mu\,\partial_\mu\phi\,,\label{Dtilde}\\
(\mathcal{\hat{D}}_\mu)^A_B
        &=&\delta^A_B\,\partial_\mu+(\Omega_\mu)^A_B\,,\label{Dhat}
\end{eqnarray}
with
\begin{eqnarray}\label{Omega}
(\Omega_\mu)^A_B=\frac{1}{4}\omega^{\hat{\nu}\hat{\rho}}_\mu\Gamma_{\hat{\nu}\hat{\rho}}\,\delta^A_B
           +\frac{i\,e^\phi}{8\cdot5!}\,\Gamma^{\kappa\nu\rho\sigma\delta}F_{\kappa\nu\rho\sigma\delta}\Gamma_\mu(\sigma_2)^A_B\,,
\end{eqnarray}
where $\sigma_2$ is the Pauli matrix,
$\omega^{\hat{\nu}\hat{\rho}}_\mu$ is the spin connection and the
hatted indices are used for the tangent space. Since
$\phi=\phi(x^+)$ , the dilatino variation imposes the constraint
\begin{eqnarray}
\Gamma^{+}\epsilon^A=0,\quad A=1,2\,. \label{DC}
\end{eqnarray}
Therefore at most the background (\ref{BG}) admits 16
supersymmetries. Now let's consider the gravitino variations. In
order to work out the spin connection
$\omega^{\hat{\nu}\hat{\rho}}_\mu$, we choose the vierbeins
$e^{\hat{\nu}}_\mu$ as follows
\begin{eqnarray}
e^{\hat{+}}_+=e^{\hat{-}}_-=1\,\,,\,e^{\hat{I}}_J=\delta^{\hat{I}}_J\,\,,\,
\,e^{\hat{-}}_+=\frac{1}{2}\lambda\,x^2_I\,\,,\label{VB}
\end{eqnarray}
It is easy to get the only non-vanishing components of
$\omega^{\hat{\nu}\hat{\rho}}_{\mu}$ are
\begin{eqnarray}
\omega^{\hat{-}\hat{I}}_+=\lambda \,x^I\,.
\end{eqnarray}
In general the Gamma matrices with respect to the coordinate, say $\Gamma^\mu,\Gamma_\mu$,
and those with respect to the vierbein, say $\Gamma^{\hat{\mu}},\Gamma_{\hat{\mu}}$,
are not the same things. They are related by $\Gamma^\mu=\Gamma^{\hat{\mu}}e_{\hat{\mu}}^\mu$ and
$\Gamma_\mu=\Gamma_{\hat{\mu}}e^{\hat{\mu}}_\mu$. In our background we have
\begin{eqnarray}
\Gamma^I=\Gamma^{\hat{I}}\,,\quad\Gamma^+=\Gamma^{\hat{+}}\,,
   \quad\Gamma_+=\Gamma_{\hat{+}}+\frac{1}{2}\lambda\, x_I^2\,\Gamma_{\hat{-}}\,,\nonumber\\
\Gamma_I=\Gamma_{\hat{I}}\,,\quad\Gamma_-=\Gamma_{\hat{-}}\,,
   \quad\Gamma^-=\Gamma^{\hat{-}}-\frac{1}{2}\lambda\, x_I^2\,\Gamma^{\hat{+}}\,.
\end{eqnarray}

Now we will consider the Killing spinor equation coming from gravitino variation
\begin{eqnarray}
(\textbf{1}\cdot\partial_\mu+\Omega_\mu)^A_B\,\epsilon^B=0\,\,,\label{GC}
\end{eqnarray}
with \begin{eqnarray}\label{CovDiff}
\Omega_-&=&0\,\,,\,\nonumber\\[0.1cm]
\Omega\,_I&=&\frac{i\,e^\phi}{4}f\,\Gamma^+(\Pi+\Pi')\,\Gamma_I\,\sigma_2\,\,,\nonumber\\
\Omega_+&=&-\frac{1}{2}\lambda\,x^I\Gamma^{+I}\,\textbf{1}+
                         \frac{i\,e^\phi}{4}f\,\Gamma^+(\Pi+\Pi')\,\Gamma_+\sigma_2\,\,,\,
\end{eqnarray}
where $\Pi=\Gamma^{1234}=\Gamma^1\Gamma^2\Gamma^3\Gamma^4$ and
$\Pi'=\Gamma^{5678}=\Gamma^5\Gamma^6\Gamma^7\Gamma^8$. Taking
(\ref{DC}) and $(\G^+)^2=0$ into account, we note that the
$\mu=-,I$ components of (\ref{GC}) are simply
\begin{eqnarray}
\partial_-\,\epsilon^A=0\,,\quad\partial_I\,\epsilon^A=0\,.
\end{eqnarray}
Therefore $\epsilon^A$ is $x^-$ and $x^I$-independent. The $\mu=+$
component of the Killing spinor equation (\ref{GC}) takes the form
\begin{eqnarray}
(\textbf{1}\cdot\partial_{+}+
    i\,e^\phi f\,\Pi\,\sigma_2)^A_B\,\,\epsilon^B=0\,\,,\label{GE}
\end{eqnarray}
where we have used the facts $\Pi \epsilon^A=\Pi' \epsilon^A$ and
$\Gamma^+\Gamma_+\epsilon^A=2\epsilon^A$ due to
$\Gamma^+\epsilon^A=0$. The equation (\ref{GE}) can be written as
\begin{eqnarray}\label{12}
\partial_{+}\epsilon^1+e^{\phi}f\,\Pi\,\epsilon^2=0,\nonumber\\
\partial_{+}\epsilon^2-e^{\phi}f\,\Pi\,\epsilon^1=0\,\,.
\end{eqnarray}
\begin{enumerate}
\item In the case of $f=f_0\,e^{-\phi}$, the equation (\ref{12})
becomes
\begin{eqnarray}
\partial_{+}\epsilon^1+f_0\,\Pi\,\epsilon^2=0,\nonumber\\
\partial_{+}\epsilon^2-f_0\,\Pi\,\epsilon^1=0\,\,.
\end{eqnarray}
The corresponding solutions are
\begin{eqnarray}
&&\epsilon^1=\chi_0\cos(f_0\,x^+)+\Pi\tilde{\chi}_0\sin(f_0\,x^+)\,,\nonumber\\[0.1cm]
&&\epsilon^2=-\tilde{\chi}_0\cos(f_0\,x^+)+\Pi\chi_0\sin(f_0\,x^+)\,;
\end{eqnarray}
\item In the case of $f=f_0$, the solutions of equation (\ref{12})
are given by
\begin{eqnarray}
&&\epsilon^1=\chi_0\cos\frac{f_0\,e^{-cx^+}}{c}+\Pi\tilde{\chi}_0\sin\frac{f_0\,e^{-cx^+}}{c}\,,\nonumber\\
&&\epsilon^2=\tilde{\chi}_0\cos\frac{f_0\,e^{-cx^+}}{c}-\Pi\chi_0\sin\frac{f_0\,e^{-cx^+}}{c}\,.
\end{eqnarray}
\end{enumerate}
Here $\chi_0$ and $\tilde{\chi}_0$ are arbitrary constant 10d
Majorana-Weyl spinors of positive  chirality and satisfy (\ref{DC}).
Therefore the background (\ref{BG}) preserves 16 supersymmetries. In
fact, the above discussion does not depend on the explicit form of
the $\lambda$ and $f$, so any supergravity solutions satisfying
(\ref{lam}) keep half of the original supersymmetries.

\subsection{Asymptotic behavior in Rosen coordinates}

The background (\ref{BG}) with linear null dilaton and constant RR
five-form flux seems to support very smooth geometry. Using the
metric in (\ref{BG}), we get the only non-vanishing component of
the Ricci tensor is $R_{++}=8\lambda$. In the case with constant
RR field strength, $\lambda=f^2_0\,e^{-2c\,x^+}$. When $x^+$ goes
from $-\infty$ to $+\infty$ , $\lambda$ decreases from $+\infty$
to zero . Taking the $x^+$ as light-cone time, the evolution is to
start from the {\it big bang} region in the past and reach the
flat space in the infinite future. In the other case with constant
$\lambda$, the evolution is from a {\it big bang} singular point
to the plane-wave spacetime in the light-cone future.  Similar to
the case studied in \cite{Verlinde}, we have a big-bang
singularity in the Einstein frame.

To see it clearly, it is better to transform to Rosen coordinates.
>From Brinkmann coordinates $(x^+,\,x^-,\,x^I)$ to Rosen
coordinates $(x^+,\,\tilde{x}^-,\,\tilde{x}^I)$ , we change the
coordinates in the following way
\begin{eqnarray}
x^-=\tilde{x}^-+\frac{1}{2}\,aa'\tilde{x}^I\tilde{x}^I\,\,,\,\,
x^I=a\tilde{x}^I\,\,,
\end{eqnarray}
with $a$ is defined in (\ref{AA}). Now the metric takes the form
\begin{eqnarray}
ds^2=-2dx^+d\tilde{x}^-+a^2(x^+)d\tilde{x}^Id\tilde{x}^I\,,
\end{eqnarray}
and thus is conformally flat. In the case with constant RR field
strength, the solution of the equation (\ref{AA}) is
\begin{eqnarray}
a(x^+)=C_1\,J_0\left(\frac{f_0}{c}\,e^{-cx^+}\right)+C_2\,Y_0\left(\frac{f_0}{c}\,e^{-cx^+}\right)\,\,,
\end{eqnarray}
where $J$ and $Y$ are standard Bessel and Neumann functions ;
$C_1$ and $C_2$ are constants. The diverse values of $(C_1,C_2)$
can be thought of different conformal embeddings of the plane wave
into Minkowski spacetime. Since $J_0$ and $Y_0$ are both
oscillating functions, $a$ oscillates with time.

Using the asymptotic expansion of Bessel functions as
$z\rightarrow+\infty$ ,
\begin{eqnarray}
J_{\nu}(z)&\sim&
       \sqrt{\frac{2}{\pi z}}\,\cos(z-\frac{\nu\pi}{2}-\frac{\pi}{4})\,\,,
            \,\,\,\,\,\,|\arg z|<\pi\,\,,\nonumber\\
Y_{\nu}(z)&\sim&
       \sqrt{\frac{2}{\pi z}}\,\sin(z-\frac{\nu\pi}{2}-\frac{\pi}{4})\,\,,\,
            \,\,\,\,\,\,|\arg z|<\pi\,\,,\label{BI}
\end{eqnarray}
we can see that as $x^+\rightarrow-\infty$
,$\frac{f_0}{c}\,e^{-cx^+}\,\rightarrow\infty$ ,
\begin{eqnarray}
&&J_0\sim e^{\frac{c}{2}x^+}
          \sqrt{\frac{2\,c}{\pi f_0}}\,\cos\left(\frac{f_0}{c}\,e^{-cx^+}\right)\rightarrow0\,\,,\nonumber\\
&&Y_0\sim e^{\frac{c}{2}x^+}
          \sqrt{\frac{2\,c}{\pi f_0}}\,\sin\left(\frac{f_0}{c}\,e^{-cx^+}\right)\rightarrow0\,\,,\nonumber\\[0.3cm]
&&a(x^+)\rightarrow0\,\,,
\end{eqnarray}
which corresponds to the big bang singularity in the Brinkmann
coordinates. Considering the asymptotic expansion of Bessel
functions as $z\rightarrow0$ ,
\begin{eqnarray}
J_{\nu}(z)&\sim&\frac{1}{\Gamma(1+\nu)}\left(\frac{z}{2}\right)^{\nu}+\mathcal{O}(z^{\nu+2})\,\,,\nonumber\\
Y_0(z)&\sim&\frac{2}{\pi}\ln\frac{z}{2}\,\,,\,\label{BZ}
\end{eqnarray}
we get as $x^+\rightarrow+\infty$ ,
$\frac{f_0}{c}\,e^{-cx^+}\hspace{-0.2cm}\rightarrow0$ ,
\begin{eqnarray}
&&J_0\sim 1\,\,\,,\,\, Y_0\sim
\frac{2}{\pi}(\,\ln\frac{f_0}{c}-cx^+)\,\,,\nonumber\\
&&a(x^+)\sim C_1+C_2\left(\,\ln\frac{f_0}{c}-cx^+\right).
\end{eqnarray}
In order to recover the flat space limit, we set $C_1=1$ and
$C_2=0$. Thus
\begin{eqnarray}
a(x^+)=J_0\left(\frac{f_0}{c}\,e^{-cx^+}\right)\,\,.
\end{eqnarray}
Therefore, in the Rosen coordinate, the metric in the string frame
looks like the Friedman-Robersen-Walker metric but now the scale
factor oscillates with the light-cone time.

\section{Bosonic sector}

It is well-known that the plane-wave backgrounds are the exact
solutions to the string theory if they solve the supergravity
equations \cite{Guven, Amati, 90Horow}. In other words, the
plane-wave background is exact against  the $\a^\prime$
correction. And in \cite{90Horow} it has been shown that the
light-cone gauge can be implemented for the string theory in a
plane-wave background. This has been generalized to the string
theory in a plane-wave background with RR field strength in
\cite{Metsaev01, MT02}.  In this  and the next sections we will
solve the first-quantized type IIB free superstring model in the
background (\ref{BG}) with linear null dilaton and constant RR
five-form flux. In the light-cone gauge the Green-Schwarz action
is quadratic in the transverse bosonic and fermionic coordinates
for the plane wave metric with any function $\lambda(x^+)$ , so we
can explicitly write down the classical solutions, perform the
canonical quantization and find the the light-cone Hamiltonian in
terms of creation and
annihilation operators. 

First we will study the bosonic sector of the model. The bosonic
part of the GS action in the background (\ref{BG}) is
\begin{eqnarray}\label{BosA}
S_B&=&-\frac{1}{4\pi \alpha'}\int
      d^2\sigma\sqrt{-g}\,g^{ab}G_{\mu\nu}\partial_{a}x^{\mu}\partial_{b}x^{\nu}\,\nonumber\\
   &=&-\frac{1}{4\pi \alpha'}\int
      d^2\sigma\sqrt{-g}\,g^{ab}(-2\,\partial_a x^{+}\partial_b x^{-}-\lambda\,
      x^2_{I}\partial_a x^{+}\partial_b x^{+}+\partial_a x^{I}\partial_b x^{I})\,.
\end{eqnarray}
First we choose the world-sheet metric as
\begin{eqnarray}
\sqrt{-g}\,g^{ab}=\left(
\begin{array}{cc}
-1 & 0 \\
0 & 1 \\
\end{array}\right)\,.\label{GG}
\end{eqnarray}
To fix the residual worldsheet gauge symmetries, we note the
equation of motion for $x^+$ is
$(\partial^2_\tau-\partial^2_\sigma)x^+=0$ , which has a general
solution of the form $f(\tau+\sigma)+g(\tau-\sigma)$ . We choose
$f(x)=g(x)=\frac{1}{2}\alpha'p^+x$ , then
\begin{eqnarray}
x^+=\alpha'p^+\tau\,\,,\,p^+>0\,.\label{BX}
\end{eqnarray}
The conditions (\ref{GG}) and (\ref{BX}) completely fix the
world-sheet gauge symmetries of the bosonic action (\ref{BosA}).
This is the well-known light-cone gauge. After imposing these gauge
choices, $x^-$ is not a dynamical variable either, which is
completely determined by $x^I$ through the constraints resulting
from
\begin{eqnarray}\label{Cg}
\frac{\delta S_B}{\delta g^{\tau\sigma}}=0\,\,,\,\,
\frac{\delta S_B}{\delta g^{\tau\tau}}=\frac{\delta S_B}{\delta g^{\sigma\sigma}}=0\,.
\end{eqnarray}
In the light-cone gauge these constrains become
\begin{eqnarray}
\partial_{\sigma}x^-&=&\frac{1}{\alpha'p^+}\partial_{\tau}x^I\partial_{\sigma}x^I\,\,,\nonumber\\
\partial_{\tau}x^-&=&\frac{1}{2\,\alpha'p^+}(\partial_{\tau}x^I\partial_{\tau}x^I+
                      \partial_{\sigma}x^I\partial_{\sigma}x^I-\lambda x^2_I)\,\,.\label{CX}
\end{eqnarray}
Without loss of generality, we set $c=\frac{1}{\alpha'p^+}$ in the
following. Then the bosonic action in the light-cone gauge takes the form
\begin{eqnarray}
S_B&=&\frac{1}{4\pi \alpha'}\int d\tau \int_0^{2\pi}\hspace{-0.2cm}d\sigma
      \,\,(\partial_{\tau}x^I\partial_{\tau}x^I-
      \partial_{\sigma}x^I\partial_{\sigma}x^I-\tilde{f}^2e^{-2\,\tau}x^2_I)\,\,,\label{BA}
\end{eqnarray}
where for simplicity we have defined $\tilde{f}\equiv\alpha'p^+f_0$
which is dimensionless.

Obviously, in the case with $f=f_0e^{-\phi}$, the action is exactly
the same as that of the maximally supersymmetric plane-wave.
Therefore the equations of motion and the canonical quantization are
also the same. We will not repeat it here. However one should keep
in mind that the dilaton is linear null rather than a constant and
the string coupling could be very strong and the perturbative string
theory is ill-defined in the strong coupling region, just like the
case studied in \cite{Verlinde}. In the remaining part of this
section, we will focus on the case with constant RR field strength.

\subsection{Equations of motion and modes expansion}
It is easy to get the equations of motions from (\ref{BA})
\begin{eqnarray}
(\partial^2_{\tau}-\partial^2_{\sigma}+\tilde{f}^2e^{-2\,\tau})\,x^I=0\,.
\end{eqnarray}
Expanding in Fourier modes in $\sigma$ , we get an infinite
collection of oscillators with time-dependent frequencies. The
general solution is given by
\begin{eqnarray}
x^I(\tau,\,\sigma)=x^I_0(\tau)+i
\sqrt{\frac{\alpha'}{2}}\sum^{\infty}_{n=1}\frac{1}{\sqrt{n}}
      \left[\,T^I_n(\tau)\,e^{in\sigma}-T^I_{-n}(\tau)\,e^{-in\sigma}\right],
      \label{Tn}
\end{eqnarray}
with
\begin{eqnarray}\label{Z_n}
x^I_0(\tau)&=&J_0(\tilde{f}e^{-\tau})\,\tilde{x}^I-\frac{\pi}{2}\,
              \alpha'\,Y_0(\tilde{f}e^{-\tau})\,\tilde{p}^I\,,\nonumber\\[0.1cm]
T^I_n(\tau)&=&Z_n(\tau)\,\alpha^I_n-Z_{-n}(\tau)\,\tilde{\alpha}^I_{-n}\,,\nonumber\\
Z_n(\tau)&=&\left(\frac{\tilde{f}}{2}\right)^{\hspace{-0.2cm}-in}\hspace{-0.2cm}
              \Gamma(1+in)\,\,J_{in}(\tilde{f}e^{-\tau})\,.
\end{eqnarray}
Notice that $Z_{-n}(\tau)=Z^{\ast}_{n}(\tau)$ and
$T^I_{-n}(\tau)=T^{I\ast}_n(\tau)$ due to the facts
$\Gamma^\ast(1+in)=\Gamma(1-in)$ and
$J^\ast_{in}(\tilde{f}e^{-\tau})=J_{-in}(\tilde{f}e^{-\tau})$.

To consider the asymptotic behavior, we need use the asymptotic
expansion of the Bessel functions as $z\rightarrow0$ in (\ref{BZ}).
We can see that as $\tau\rightarrow+\infty$ ,
\begin{eqnarray}
x^I_0(\tau)&\thicksim&\tilde{x}^I+\alpha'(\tau-\ln\frac{\tilde{f}}{2})\,\tilde{p}^I\,,\nonumber\\
Z_{n}(\tau)&\thicksim&e^{-in\tau}\,.
\end{eqnarray}
We come to the flat-space theory just as expected. Note that in our
case the asymptotic flatness behavior is also shared by the zero
modes, which is different from the paper \cite{02Tsey}. Requiring
that $x^I$ are real functions implies that
\begin{eqnarray}
(\alpha^I_n)^\dag=\alpha^I_{-n}\,\,,\,\,(\tilde{\alpha}^I_n)^\dag=\tilde{\alpha}^I_{-n}\,.
\end{eqnarray}
The canonical momenta $\Pi^I$ and the total momentum carried by the
string are
\begin{eqnarray}
\Pi^I=\frac{1}{2\pi\alpha'}\,\partial_{\tau}x^I\,\,,\,\,\,\,
\,p^I_0=\int^{2\pi}_0d\sigma\,\Pi^I=\frac{1}{\alpha'}\,\partial_{\tau}x^I_0\,.
\end{eqnarray}
To quantize the theory, we need to impose the canonical commutation
relations
\begin{eqnarray}
[\,x^I(\tau,\sigma),\,\Pi^J(\tau,\sigma')\,]=i\,\delta^{IJ}\delta(\sigma-\sigma')\,,\label{CR}
\end{eqnarray}
with the other commutators vanishing. These commutators are ensured
by requiring the following commutators for the modes
\begin{eqnarray}
[\tilde{x}^I,\,\tilde{p}^J]=i\,\delta^{IJ}\,\,,\,\,\,\,
[\alpha^I_n,\,\alpha^{J\dag}_m]=\delta^{IJ}\delta_{nm}\,\,,\,\,\,\,
[\tilde{\alpha}^I_n,\,\tilde{\alpha}^{J\dag}_m]=\delta^{IJ}\delta_{nm}\,\,.\label{OC}
\end{eqnarray}
>From this we can also get
\begin{eqnarray}
[\,x^I_0(\tau),\,p^J_0(\tau)\,]=i\,\delta^{IJ}\,.
\end{eqnarray}
Note that in these calculations we have used the formulae
\begin{eqnarray}\label{Bes}
&&\Gamma(1+in)\,\Gamma(1-in)=\frac{n\pi}{\,\sinh{n\pi}}\,\,,\nonumber\\
&&J_{\nu}(z)J'_{-\nu}(z)-J_{-\nu}(z)J'_{\nu}(z)=-\frac{2\sin{\nu\pi}}{\pi z}\,\,.
\end{eqnarray}

\subsection{Light-cone Hamiltonian}

According to (\ref{BA}) , the bosonic part of the light-cone
Hamiltonian of our model is
\begin{eqnarray}
H_B=\frac{1}{4\pi\alpha'^2p^+}\int^{2\pi}_0d\sigma
      (\partial_{\tau}x^I\partial_{\tau}x^I+\partial_{\sigma}x^I\partial_{\sigma}x^I
      +\tilde{f}^2e^{-2\,\tau}x^2_I)\,\,.
\end{eqnarray}
Inserting the mode expansion of $x^I$ in terms of creation and
annihilation operators, we obtain
\begin{eqnarray}\label{AC}
H_B&=&H_{B0}(\tau)+\frac{1}{2\alpha'p^+}
   \sum^{\infty}_{n=1}\left[\,\Omega^B_n(\tau)\,
   (\,\alpha^{I\dag}_{n}\alpha^I_n+\tilde{\alpha}^{I\dag}_{n}\tilde{\alpha}^I_n+1)\right.\nonumber\\
&&\hspace{4cm}\left.-C^B_n(\tau)\,\alpha^I_{n}\tilde{\alpha}^I_n
   -C^{B\ast}_n(\tau)\,\tilde{\alpha}^{I\dag}_{n}\alpha^{I\dag}_{n}\,\right]\,,
\end{eqnarray}
with
\begin{eqnarray}
H_{B0}(\tau)&=&
    \frac{1}{2p^+}\left[(p^I_0)^2+\tilde{f}^2e^{-2\tau}\left(\frac{x^I_0}{\alpha'}\right)^2\right]\,\,,\nonumber\\
\Omega^B_n(\tau)&=&
   \frac{1}{n}|\partial_{\tau}Z_n|^{2}+n\left(1+\frac{\tilde{f}^2e^{-2\tau}}{n^{2}}\right)|Z_n|^{2}\,\,,\nonumber\\
C^B_n(\tau)&=&
    \frac{1}{n}(\partial_{\tau}Z_n)^{2}+n\left(1+\frac{\tilde{f}^2e^{-2\tau}}{n^{2}}\right)(Z_n)^{2}\,\,.
\end{eqnarray}
Now let's see how the functions $\Omega^B_n(\tau)$ and $C^B_n(\tau)$
behave at infinite $\tau$ . Taking into account the asymptotic
expansion of Bessel functions (\ref{BZ}) , we obtain that as
$\tau\rightarrow+\infty$,
\begin{eqnarray}
\Omega^B_n(\tau)\sim2n+\mathcal{O}(e^{-\tau})\,\,\,,\,\,\,C^B_n(\tau)\sim0+\mathcal{O}(e^{-\tau})\,\,,
\end{eqnarray}
where $C^B_n(\tau)$ is completely suppressed. We get the result just
as in the flat space. On the other hand, when $\tau$ goes to
$-\infty$ , we come to the strongly coupled region with
\begin{eqnarray}
\Omega^B_n(\tau)&\sim&\frac{2\cosh {n\pi}}{\sinh {n\pi}}\,\tilde{f}e^{-\tau}\,,\nonumber\\[0.2cm]
C^B_n(\tau)&\sim&2\left(\frac{\tilde{f}}{2}\right)^{\hspace{-0.2cm}-2in}\hspace{-0.1cm}
        \frac{\Gamma^2(1+in)}{\pi}\,\tilde{f}e^{-\tau}\,,
\end{eqnarray}
where we have used (\ref{BI}). Now $\Omega^B_n(\tau)$ and $C^B_n(\tau)$
are of the same order.

It is evident that the Hamiltonian (\ref{AC}) is non-diagonal. The
treatment of the zero-mode part is the standard way used in
point-particle quantization,see \cite{02Tsey}. The non-zero mode
part has non-diagonal terms proportional to $C^B_n(\tau)$ and
$C^{B\ast}_n(\tau)$ . The evolution of generic states made out of
$\alpha^I_n$ , $\tilde{\alpha}^I_n$ is thus non-trivial. In the
following we will find a new set of modes to make  the Hamiltonian
diagonal. Now let us introduce a new set of time-dependent string
modes $A^I_n$ , $\tilde{A}^I_n$ which are defined by
\begin{eqnarray}
\frac{i}{\sqrt{|n|}}\left[Z_n\alpha^I_n-Z^\ast_n\tilde{\alpha}^I_{-n}\right]&=&
           \frac{i}{\sqrt{|\,\omega_n|}}\left[e^{-i\omega_n\tau}A^I_n(\tau)-
           e^{i\omega_n\tau}\tilde{A}^I_{-n}(\tau)\right]\,\,,\nonumber\\
\frac{i}{\sqrt{|n|}}\left[\partial_{\tau}Z_n\,\alpha^I_n-\partial_{\tau}Z^\ast_n\,\tilde{\alpha}^I_{-n}\right]&=&
           \sqrt{|\,\omega_n|}\left[e^{-i\omega_n\tau}A^I_n(\tau)+
           e^{i\omega_n\tau}\tilde{A}^I_{-n}(\tau)\right]\,\,\,,
\end{eqnarray}
where
\begin{eqnarray}
\omega_n=\sqrt{n^2+\tilde{f}^2e^{-2\tau}}\,\,,\,n>0\quad;\quad
\omega_{-n}=-\sqrt{n^2+\tilde{f}^2e^{-2\tau}}\,\,,\,n<0\,,
\end{eqnarray}
with similar relations $A^{\dag}_n=A_{-n}$ and
$\tilde{A}^{\dag}_n=\tilde{A}_{-n}$ . It follows then
\begin{eqnarray}\label{DA}
A^I_n(\tau)&=&\alpha^I_n\,f_n(\tau)+\tilde{\alpha}^I_{-n}\,g^{\ast}_n(\tau)\,\,,\,\,\,\,
A^{I\dag}_n(\tau)\,\,=\,\,\alpha^I_{-n}\,f^{\ast}_n(\tau)+\tilde{\alpha}^I_{n}\,g_n(\tau)\,\,,\,\,\,\,\nonumber\\
\tilde{A}^I_n(\tau)&=&\alpha^I_{-n}\,g^{\ast}_n(\tau)+\tilde{\alpha}^I_{n}\,f_n(\tau)\,\,,\,\,\,\,
\tilde{A}^{I\dag}_n(\tau)\,\,=\,\,\alpha^I_{n}\,g_n(\tau)+\tilde{\alpha}^I_{-n}\,f^{\ast}_n(\tau)\,\,,
\end{eqnarray}
where
\begin{eqnarray}\label{fg}
f_n(\tau)&=&\frac{1}{2}\,\sqrt{\frac{\omega_n}{n}}\,\,e^{i\omega_n\tau}\left[Z_n+\frac{i}{\omega_n}\partial_{\tau}Z_n\right]\,\,,\nonumber\\
g_n(\tau)&=&\frac{1}{2}\,\sqrt{\frac{\omega_n}{n}}\,\,e^{-i\omega_n\tau}\left[-Z_n+\frac{i}{\omega_n}\partial_{\tau}Z_n\right]\,\,,\
\end{eqnarray}
After some calculations using the commutation relations (\ref{OC})
and the properties (\ref{Bes}) of the Bessel functions, we obtain
the non-vanishing commutators:
\begin{eqnarray}
[A^I_n\,\,,\,A^{J\dag}_m]=\delta_{nm}\,\delta^{IJ}\,\,,\,\,\,\,
[\tilde{A}^I_n\,\,,\,\tilde{A}^{J\dag}_m]=\delta_{nm}\,\delta^{IJ}\,\,.\nonumber\\
\end{eqnarray}
In terms of these new operators, the mode expansion of
$x^I(\tau,\,\sigma)$ takes the form
\begin{eqnarray}
x^I(\tau,\,\sigma)=x^I_0(\tau)+i\sqrt{\frac{\alpha'}{2}}\sum^{\infty}_{n=1}\frac{1}{\sqrt{\omega_n}}
     \left[\,e^{-i\omega_n\tau}\left(A^I_n(\tau)\,e^{in\sigma}
     +\tilde{A}^I_n(\tau)\,e^{-in\sigma}\right)\right.\nonumber\\
\left.-\,e^{i\omega_n\tau}\left(A^I_{-n}(\tau)\,e^{-in\sigma}+\tilde{A}^{I}_{-n}(\tau)\,e^{in\sigma}\right)
    \right]\,,
\end{eqnarray}
and the bosonic Hamiltonian is
\begin{eqnarray}\label{[A]}
H_B=H_{B0}(\tau)+\frac{1}{\alpha'p^+}\sum^{\infty}_{n=1}\omega_n(\tau)\left[A^I_{-n}(\tau)A^I_n(\tau)
         +\tilde{A}^I_{-n}(\tau)\tilde{A}^I_n(\tau)+1\right],
\end{eqnarray}
which is diagonal.  The bosonic Hamiltonian reminisces the 2-d field
theory of free scalars with time-dependent masses.

One may formally define the time-dependent vacuum by
 \be
 A^I_n |0, \tau\rangle=0, \hspace{3ex} \tilde{A}^I_n|0,\tau
 \rangle =0, \hspace{3ex} n>0\,,
 \ee
 and the excitation spectrum can be obtained straightforwardly.
 The spectrum is also time-dependent.

\subsection{Classical evolution of a rotating string}

To shed some light on the dynamics of the strings in our
background, let us consider the behavior of the classical
solutions. As an simple example, we focus on the  rotating string.

First recall that in flat space, a rigid string rotating in a plane
is described by a state on the leading Regge trajectory with maximal
momentum for a given energy. In the light-cone gauge, the
corresponding solution is
\begin{eqnarray}\label{CC}
x^+&=&\alpha'p^+\tau\,\,,\,\,\,\,x^-\,=\,\alpha'p^-\tau\,\,,\,\,\,\,L^2\,=\,2\,\alpha'p^+p^-\,\,,\nonumber\\
\textsc{x}&=&x^1+ix^2\,=\,L\,e^{-i\,\tau}\cos\sigma\,\,,
\end{eqnarray}
where $x^1$ and $x^2$ are Cartesian coordinates of the transverse
2-plane. Here we have the standard leading Regge trajectory relation
\begin{eqnarray}\label{Regge}
-2\,\alpha'p^+p^-=\alpha'(E-p^2_y)=2\,J\,\,,
\end{eqnarray}
with $p_y$ representing the momentum in the transverse directions
and $J$ standing for the corresponding angular momentum.

Now let us turn to our time-dependent background (\ref{BG}). In the
case with constant $\lambda$, the analogue of the rotating string
solution is given by
\begin{eqnarray}
x^+&=&\alpha'p^+\tau\,\,,\nonumber\\
\textsc{x}&=&x^1+ix^2\,=\,L\,e^{-i\omega\tau}\cos\sigma\,\,,
\end{eqnarray}
with $\omega=\sqrt{1+(\alpha'p^+f^2_0)}$ and $x^-$ determined by the
constraints (\ref{Cg}). After some calculations, we get the
light-cone Hamiltonian
\begin{eqnarray}\label{H1}
H=-p_+=p^-=\frac{L^2}{4\,\alpha'^2p^+}\,\omega^2\,\,,
\end{eqnarray}
and the angular momentum associated with rotations in the transverse
space
\begin{eqnarray}\label{J1}
J=\frac{1}{2\pi\alpha'}\int^{2\pi}_0d\sigma(x^1\dot{x}^2-\dot{x}^1x^2)=-\frac{L^2}{2\alpha'}\,\omega\,\,.
\end{eqnarray}
Combining (\ref{H1}) and (\ref{J1}), we find the direct analogue of
the flat-space Regge trajectory relation:
\begin{eqnarray}
-2\,\alpha'p^+p^-=\omega\cdot2J\,,
\end{eqnarray}
with the ``effective tension'' $T=\frac{\omega}{2\pi\alpha'}$
which is still time-independent. Just as expected, this result is
similar to the maximally supersymmetric case.

In the case with constant RR field strength, the analogue of the
rotating string solution takes the form
\begin{eqnarray}\label{QC}
x^+&=&\alpha'p^+\tau\,\,,\nonumber\\
\textsc{x}&=&x^1+ix^2\,=\,L\,Z_1(\tau)\cos\sigma\,\,,
\end{eqnarray}
and $x^-$ is again determined by the constraints (\ref{Cg}) and
thus evolves with time in a non-trivial way. $Z_1(\tau)$ is
defined in (\ref{Z_n}). It is easy to check that as
$\tau\rightarrow+\infty$ , we recover the flat space result
(\ref{CC}). In general, it describes a rotating string with
effective length $L_{eff}=L\,Z_1(\tau)$ which oscillates with time
and rapidly shrinks to zero as $\tau\rightarrow-\infty$ .

Inserting (\ref{QC}) into the Hamiltonian (\ref{BA}) , we can get
the light-cone energy
\begin{eqnarray}\label{H2}
H=-p_+=p^-=\frac{L^2}{4\,\alpha'^2p^+}\Omega^B_1(\tau)\,\,.
\end{eqnarray}
The angular momentum associated with rotations in the transverse
space is
\begin{eqnarray}\label{J2}
J=-\frac{L^2}{2\,\alpha'}\,\,.
\end{eqnarray}
Combining (\ref{H1}) and (\ref{J2}) , it is interesting to note that
we find
\begin{eqnarray}
-2\,\alpha'p^+p^-=\Omega^B_1(\tau)\,J\,\,,
\end{eqnarray}
which is analogous to the standard leading Regge trajectory relation
(\ref{Regge}) but with an ``effective tension'' function
$T=\frac{1}{2\pi\alpha'}\cdot\frac{1}{2}\Omega^B_1(\tau)$ . At
$\tau\rightarrow+\infty$ , we have $\Omega^B_1\rightarrow2$ . Thus we
reach the flat space limit. As we go back in time, the energy of
this state rapidly grows until it diverges as
$\tau\rightarrow-\infty$.

\section{Fermionic sector}

For a general plane-wave background the light-cone gauge
Green-Schwarz action  comes from \cite{MT02, Cvetic1, Cvetic2, 02Russo}
\begin{eqnarray}\label{FerA}
S_F=-\frac{i}{2\pi\alpha'}\int d^2\sigma
    (\sqrt{-g}g^{ab}\delta_{AB}-\epsilon^{ab}\sigma_{3AB})\,\partial_{a}x^{\mu}\,
    \bar{\theta}^A\Gamma_{\mu}(\hat{D}_b\theta)^B\,.
\end{eqnarray}
Here $\sigma_3=\mathrm{diag}(1,-1)$ and $\hat{D}_b=\partial_b+\Omega_\nu\,\partial_{b}x^\nu$ is
the pull-back of the generalized covariant
derivative $\hat{\mathcal{D}}_\nu$ (\ref{Dhat}) to the world-sheet
with $\Omega_\nu$ defined in (\ref{CovDiff}) in our background.
In (\ref{FerA}) two fermionic coordinates $\theta^{A=1,2}$ are
10d Majorana-Weyl spinors. We choose the representation of $\Gamma$-matrices
such that $\Gamma^0=C$, with $C$ being the 10d charge conjugation,
so the components of $\theta^{A}$ are all real.
For more about our convention see Appendix A.

To fix the world-sheet gauge symmetries, in addition to the bosonic
light-cone gauge conditions
$\sqrt{-g}g^{ab}=\eta^{ab}=\mathrm{diag}(-1,1)$ and
$x^+=\alpha'p^+\tau$, we need to impose
\begin{eqnarray}\label{Gamma+}
\Gamma^{+}\theta^{A}=0\,,
\end{eqnarray}
since the fermionic action (\ref{FerA}) has an additional local
$\kappa$-symmetry. These three conditions completely fix the
world-sheet gauge symmetries of the fermionic action (\ref{FerA}).
Note that due to the gauge condition (\ref{Gamma+}) only the
second term in $\Omega_+$ has non-vanishing contributions. After
some calculations, the fermionic action in the light-cone gauge
can be written as
\begin{eqnarray}\label{FA light-cone}
S_F=\frac{\,\,i\,p^+}{\sqrt{2}\,\pi}\int
d\tau\int_0^{2\pi}\hspace{-0.2cm}d\sigma\,
    (\theta^{1T}\partial_\tau\theta^1+\theta^{2T}\partial_\tau\theta^2
    +\theta^{1T}\partial_\sigma\theta^1-\theta^{2T}\partial_\sigma\theta^2
    +2\tilde{f}e^{-\tau}\theta^{1T}\Pi\theta^2)\,.
\end{eqnarray}

Similar to the bosonic case, when $f=f_0e^{-\phi}$, the action
reduces to the one in the usual plane-wave metric. The
quantization procedure follows straightforwardly. In the following
we focus on the case with constant RR field strength.

\subsection{Equations of motion and modes expansion}

>From the light-cone action (\ref{FA light-cone}), it is easy to
get the equation of motion of the fermionic sector
\begin{eqnarray}\label{1-ord}
&&(\,\partial_\tau + \partial_\sigma)\,\theta^1
        +\tilde{f}e^{-\tau}\Pi\theta^2=0\,\,,\nonumber\\
&&(\,\partial_\tau - \partial_\sigma)\,\theta^2
        -\tilde{f}e^{-\tau}\Pi\theta^1=0\,.
\end{eqnarray}
This is a system of two coupled first order equations, from which we
can obtain two decoupled second order equations as
\begin{eqnarray}\label{2-ord}
&&(\partial_\tau^2-\partial_\sigma^2)\,\theta^1+(\partial_\tau+\partial_\sigma)\,\theta^1
  +\tilde{f}^2e^{-2\tau}\theta^1=0\,\,,\nonumber\\
&&(\partial_\tau^2-\partial_\sigma^2)\,\theta^2+(\partial_\tau-\partial_\sigma)\,\theta^2
  +\tilde{f}^2e^{-2\tau}\theta^2=0\,\,.
\end{eqnarray}
Note that the time-dependent dilaton and the nonzero RR field
strength play an important role. That is, in these two decoupled
second order equations, first order differential terms appear, which
do not exist in neither the maximally supersymmetric plane-wave
background \cite{Blau,Metsaev01,MT02} due to its time independence, nor
in the time-dependent plane-wave model studied in \cite{02Tsey} because
of the absence of the RR fluxes. The solutions of (\ref{1-ord}) are\footnote
{Actually $\theta^A$ and
$\beta_n\,,\,\tilde{\beta}_n$ etc. are all 10d spinors. If we write
down the spinor indices explicitly then they should be
$\theta^A_\alpha$ and $\beta_{n\alpha}\,,\,\tilde{\beta}_{n\alpha}$
etc. with $\alpha=1,2,\ldots,32\,$. Something like $\beta_n^\dag$
and $\tilde{\beta}_n^\dag$ should be understood as
$\beta_{n\alpha}^\dag$ and $\tilde{\beta}_{n\alpha}^\dag$.}
\begin{eqnarray}\label{mode}
\theta^1(\tau,\sigma)&=&\theta^1_0(\tau)
   +\sum_{n=1}^\infty \left[\,\theta_n^1(\tau)\,e^{in\sigma}+\theta_{-n}^1(\tau)\,e^{-in\sigma}\right]\,\,,\nonumber\\
\theta^2(\tau,\sigma)&=&\theta^2_0(\tau)
   +\sum_{n=1}^\infty \left[\,\theta_n^2(\tau)\,e^{in\sigma}+\theta_{-n}^2(\tau)\,e^{-in\sigma}\right]\,\,,
\end{eqnarray}
with
\begin{eqnarray}\label{z.m}
\theta^1_0(\tau)=\frac{1}{\sqrt{4\pi\xi}}\,(\,\beta_0\cos{u}+\Pi\tilde{\beta}_0\sin{u})\,,\nonumber\\
\theta^2_0(\tau)=\frac{1}{\sqrt{4\pi\xi}}\,(\,\tilde{\beta}_0\cos{u}-\Pi\beta_0\sin{u})\,,\label{z.m.2}
\end{eqnarray}
and
\begin{eqnarray}
\theta_n^1(\tau)=\frac{1}{\sqrt{4\pi\xi}}
  \left(\beta_{n}W_n(\tau)+\Pi\tilde{\beta}_{-n}\,\tilde{W}^\ast_{n}(\tau)\right)\,\,,\label{T^1}\\
\theta_n^2(\tau)=\frac{1}{\sqrt{4\pi\xi}}
  \left(\tilde{\beta}_{-n}\,W^\ast_{n}(\tau)-\Pi\beta_{n}\tilde{W}_n(\tau)\right)\,\,,\label{T^2}
\end{eqnarray}
where for simplicity we have defined
$u=\tilde{f}e^{-\tau},\,\xi=\frac{p^+}{\sqrt{2}\,\pi}$ and
\begin{eqnarray}
W_n(\tau)&=&\left(\frac{\tilde{f}}{2}\right)^{\hspace{-0.2cm}-in}
    \hspace{-0.1cm}\Gamma(\frac{1}{2}+in)\sqrt{\frac{u}{2}}\,J_{-\frac{1}{2}+in}(u)\,,\\
\tilde{W}_n(\tau)&=&\left(\frac{\tilde{f}}{2}\right)^{\hspace{-0.2cm}-in}
    \hspace{-0.1cm}\Gamma(\frac{1}{2}+in)\sqrt{\frac{u}{2}}\,J_{\frac{1}{2}+in}(u)\,.
\end{eqnarray}
The requirement that $\theta^{1,2}$ are real implies
\begin{eqnarray}
\beta_{-n}=\beta^{\dag}_{n}\,\,,\quad
\tilde{\beta}_{-n}=\tilde{\beta}^{\dag}_{n}\,\,,\quad n=0,\pm1,\pm2,\ldots
\end{eqnarray}
According to the action (\ref{FA light-cone}), the canonical
momentum $\mathcal{P}^A_\alpha$ conjugate to the fermionic
coordinates $\theta^A_\alpha$ are defined as (we have written down
the spinor indices explicitly)
\begin{eqnarray}
\mathcal{P}^A_\alpha=\frac{i\,p^+}{\sqrt{2}\,\pi}\,\theta^A_\alpha\,,
\quad A=1,2\quad\mathrm{and}\quad\alpha,\beta=1,2,\ldots,32\,.
\end{eqnarray}
To quantize the theory, we impose the standard anticommutation relations
\begin{eqnarray}\label{FAC}
\{\theta^A_{n\alpha}(\tau,\sigma)\,,\theta^B_{m\beta}(\tau,\sigma')\}
   =\frac{1}{2\,\xi}\,\delta^{AB}\,\delta_{n+m,0}\,\delta_{\alpha\beta}\,\delta(\sigma-\sigma'),
\end{eqnarray}
with the other anticommutators vanishing. These anticommutators are
ensured by the following relations of the fermionic creation and
annihilation operators
\begin{eqnarray}
\{\beta_{n\alpha},\beta_{m\beta}\}
=\{\tilde{\beta}_{n\alpha},\tilde{\beta}_{m\beta}\}
=\delta_{\alpha\beta}\,\delta_{n+m,0}\,,\quad n,m=0,\pm1,\pm2,\ldots
\end{eqnarray}
with the other anticommutators all becoming zero. Here we have used
the formulae
\begin{eqnarray}
&&\Gamma(\frac{1}{2}+in)\,\Gamma(\frac{1}{2}-in)=\frac{\pi}{\,\cosh{n\pi}}\,\,,\label{GaHalf}\\
&&J_{-\frac{1}{2}+in}(z)J_{-\frac{1}{2}-in}(z)+J_{\frac{1}{2}+in}(z)J_{\frac{1}{2}-in}(z)
                       =\frac{2\cosh{n\pi}}{\pi z}\,\,.\label{J-Half}
\end{eqnarray}

Now let's see whether we can recover the flat-space result when
$\tau$ goes to positive infinity. Using the asymptotic expansion
(\ref{BZ}) of the Bessel functions , we can see that as
$\tau\rightarrow+\infty$ ,
\begin{eqnarray}
\theta^1_0(\tau)\sim\frac{1}{\sqrt{4\pi\xi}}\,\,\beta_0\,\,\,\,,\,\,\,
\theta^2_0(\tau)\sim\frac{1}{\sqrt{4\pi\xi}}\,\,\tilde{\beta}_0\,\,.
\end{eqnarray}
Therefore, just like the bosonic case, the asymptotic ``flatness''
behavior is also shared by the zero modes. For the oscillating
modes,
\begin{eqnarray}
W_n(\tau)&\sim& e^{-in\tau}\,\,,\nonumber\\
\tilde{W}_n(\tau)&\sim&0\,\,,
\end{eqnarray}
where the second term vanishes due to the factor $e^{-\tau}$. So,
for $n\neq0$ , we get the standard left and right moving plane
wave. But it seems that we have lost half degrees of freedom.
However this is not the case. Note that at finite $\tau$, the
constraint of the coupled 1st order equations (\ref{1-ord}) makes
$\theta^1$ and $\theta^2$ not independent. The degrees of freedom
are presented by $\beta$ and $\tilde{\beta}$. When $\tau$ goes to
positive infinity, $\theta^1$ and $\theta^2$ decouple. Therefore
$\theta^1$ and $\theta^2$ both contribute to the total degrees of
freedom which are still denoted by $\beta$ and $\tilde{\beta}$. So
we are happy to see that we have not lost any degree of freedom
when we go to the flat space limit. Next we will show that the
total number of degree of freedom in the fermionic sector is equal
to that of the bosonic case. Recall that $\theta^A$ are ten
dimensional Majorana-Weyl fermions with 16 independent real
components. The constraint from fixing $\kappa$-symmetry by
$\Gamma^+\theta=0$ leaves us with each $\theta$ having 8
independent components. So the total degree of freedom is 16, just
as in the bosonic case (eight $x^I$'s with left and right moving
modes).

\subsection{Light-cone Hamiltonian}

According to (\ref{FA light-cone}),  the fermionic light-cone
Hamiltonian of our model is given by
\begin{eqnarray}
H_F&=&-\frac{i}{\sqrt{2}\,\pi\alpha'}\int^{2\pi}_{0}\hspace{-0.2cm}d\sigma
       \left(\theta^{1T}\partial_\sigma\theta^1-\theta^{2T}\partial_\sigma\theta^2
       +2\tilde{f}e^{-\tau}\theta^{1T}\Pi\theta^2\right)\,\nonumber\\[0.1cm]
  &=&\frac{i}{\sqrt{2}\,\pi\alpha'}\int^{2\pi}_{0}\hspace{-0.2cm}d\sigma
      \left(\,\theta^{1T}\partial_\tau\theta^1+\theta^{2T}\partial_\tau\theta^2\right)\,\,,
\end{eqnarray}
In the second line we have used the equation of motion
(\ref{1-ord}) to simplify the expression. Inserting the mode
expansion of $\theta^A$, we rewrite the Hamiltonian in terms of
creation and annihilation operators as
\begin{eqnarray}
H_F&=&H_{F0}(\tau)+\frac{1}{2\alpha'p^+}
  \sum^{\infty}_{n=1}\left[\,\Omega^{F}_n(\tau)\,
  (\,\beta^\dag_{n}\beta_n+\tilde{\beta}^\dag_{n}\tilde{\beta}_n-1)\right.\nonumber\\
&&\hspace{4cm}\left.-C^{F}_{n}(\tau)\,\beta_n\Pi\tilde{\beta}_n
  -C^{F\ast}_{n}(\tau)\,\tilde{\beta}^\dag_{n}\Pi\beta^\dag_{n}\,\right]\,,
\end{eqnarray}
with
\begin{eqnarray}
H_{F0}(\tau)&=&-\frac{2\,i}{\alpha'p^+}\,\tilde{f}\,e^{-\tau}\beta_0\Pi\tilde{\beta}_0\,\,,\label{HF0}\\
\Omega^{F}_n(\tau)&=&-2\,i\left[W_n\partial_{\tau}W^{\ast}_n+\tilde{W}_n\partial_{\tau}\tilde{W}^{\ast}_n\right]\,\,,\label{Omega^F}\\[0.1cm]
C^F_n(\tau)&=&-2\,i\left[W_n\partial_{\tau}\tilde{W}_n-\tilde{W}_n\partial_{\tau}W_n\,\right]\,\,.\label{C^F}
\end{eqnarray}
Here although the function $\Omega^{F}_n$ is formally complex, we
can prove that it is not only real but also positive definitely (see
Appendix B), just as being expected. Now let's see the behavior of
functions $\Omega^{F}_n(\tau)$ and $C^F_n(\tau)$ at infinite $\tau$.
Using the asymptotic expansion of Bessel functions (\ref{BZ}), we
obtain as $\tau\rightarrow+\infty$,
\begin{eqnarray}
\Omega^{F}_n(\tau)\sim2n\quad\,,\quad C^F_n(\tau)\sim0\,\,.
\end{eqnarray}
Therefore we recover the flat space result.  Considering the
asymptotic expansion of Bessel functions (\ref{BI}), we get as
$\tau\rightarrow-\infty$,
\begin{eqnarray}
\Omega^{F}_n(\tau)&\sim&\frac{2\sinh{n\pi}}{\cosh{n\pi}}\tilde{f}\,e^{-\tau}\,\,,\nonumber\\
C^F_n(\tau)&\sim&2\,i\left(\frac{\tilde{f}}{2}\right)^{\hspace{-0.2cm}-2in}\hspace{-0.1cm}
           \frac{\Gamma^2(\frac{1}{2}+in)}{\pi}\tilde{f}\,e^{-\tau}\,\,.
\end{eqnarray}
$\Omega^{F}_n(\tau)$ and $C^F_n(\tau)$ are of the same order in
the strongly coupled region. In the following we will make the
Hamiltonian diagonal.

Comparing with the diagonalization of the bosonic Hamiltonian, the
diagonalization of  the fermionic Hamiltonian is more tricky. We
define a group of new time-dependent creation and annihilation
operators, $B_n(\tau),\,\tilde{B}_n(\tau)$ and their conjugate, by
the following Bogolioubov-type transformation
\begin{eqnarray}
B_n=\cos\varphi_n\,\,\beta_n-ie^{i\psi_n}\sin\varphi_n\,\,\Pi\tilde{\beta}_n^\dag\,,&\quad
B_n^\dag=\cos\varphi_n\,\,\beta_n^\dag+ie^{-i\psi_n}\sin\varphi_n\,\,\Pi\tilde{\beta}_n\,,\label{B}\\
\tilde{B}_n=\cos\varphi_n\,\,\tilde{\beta}_n+ie^{i\psi_n}\sin\varphi_n\,\,\Pi\beta_n^\dag\,,&\quad
\tilde{B}_n^\dag=\cos\varphi_n\,\,\tilde{\beta}_n^\dag-ie^{-i\psi_n}\sin\varphi_n\,\,\Pi\beta_n\,.\label{tB}
\end{eqnarray}
Here $\varphi_n$ and $\psi_n$ are real functions of $\tau$ to be
determined by the requirement that the fermionic Hamiltonian is
diagonal. In (\ref{B}) and (\ref{tB}) they are restricted to
$n\geq1$. For negative $n$ we define that $B_{-n}=B_n^\dag$ and
$\tilde{B}_{-n}=\tilde{B}_n^\dag$, which implies that
$\varphi_{-n}=-\varphi_n$ and $\psi_{-n}=-\psi_n$ . The operators
$B_n,\tilde{B}_n$ defined in this way automatically satisfy the
standard anticommutative relations\footnote {In these
anticommutators we write down the spinor indices $\alpha,\beta$
explicitly.}
\begin{eqnarray}
\{B_{n\alpha},B_{m\beta}^\dag\}
=\{\tilde{B}_{n\alpha},\tilde{B}_{m\beta}^\dag\}=\delta_{nm}\,\delta_{\alpha\beta}\,,
\end{eqnarray}
with the other anticommutators vanishing due to $\Pi^2=1$ and
$\cos^2\varphi+\sin^2\varphi=1$.

In terms of these new time-dependent creation and annihilation
operators the fermionic Hamiltonian (\ref{FA light-cone}) can be
written as
\begin{eqnarray}\label{FH-Diag1}
H_F&=&H_{F0}(\tau)+\frac{1}{\alpha'p^+}\sum_{n=1}^\infty
    \left[\,\tilde{\omega}_n(\tau)\left(B_n^\dag(\tau)B_n(\tau)
    +\tilde{B}_n^\dag(\tau)\tilde{B}_n(\tau)-1\right)\right.\nonumber\\
&&\hspace{3.5cm}\left.+\,K(\tau)\,B_n(\tau)\Pi\tilde{B}_n(\tau)
    +K^\ast(\tau)\,\tilde{B}_n^\dag(\tau)\Pi B_n^\dag(\tau)\,\right].
\end{eqnarray}
Here the functions $\tilde{\omega}_n\,,K_n$ are
\begin{eqnarray}
\tilde{\omega}_n&=&\frac{1}{2}\,\Omega_n^F\cos2\varphi_n
    -\frac{i}{4}\,e^{i\psi_n}C_n^F\sin2\varphi_n+\frac{i}{4}\,e^{-i\psi_n}C_n^{F\ast}\sin2\varphi_n\,,\\
K&=&\frac{i}{2}\,e^{-i\psi_n}\Omega_n^F\sin2\varphi_n
    -\frac{1}{2}C_n^F\cos^2\varphi_n-\frac{1}{2}\,e^{-2i\psi_n}C_n^{F\ast}\sin^2\varphi_n\,.
\end{eqnarray}
with $\Omega_n^F$ and $C_n^F$ defined in (\ref{Omega^F}) and
(\ref{C^F}). To diagonalize the Hamiltonian we should require the
coefficient $K_n$ of the non-diagonal terms to vanish, i.e.
\begin{eqnarray}\label{K=0}
i\,\Omega_n^F\sin2\varphi_n=e^{i\psi_n}C_n^F\cos^2\varphi_n+e^{-i\psi_n}C_n^{F\ast}\sin^2\varphi_n\,.
\end{eqnarray}
We can always choose $\psi_n$ such that
\begin{eqnarray}
e^{i\psi_n}C_n^F=i|C_n^F| \quad\mathrm{and}\quad
e^{-i\psi_n}C_n^{F\ast}=-i|C_n^F|\,,\label{psi_n}
\end{eqnarray}
then we get the expression of the unknown function $\varphi_n$ from
(\ref{K=0}) as following
\begin{eqnarray}
\varphi_n(\tau)=\frac{1}{2}\arctan\frac{\,\,|C_n^F(\tau)|\,}{\Omega_n^F(\tau)}\,\,.\label{phi_n}
\end{eqnarray}
Now we have determined the coefficients in the definition (\ref{B})
(\ref{tB}) of $B_n$ and $\tilde{B}_n$. The last step is to calculate
the function $\tilde{\omega}_n$ in front of the diagonal terms in
(\ref{FH-Diag1}) by using (\ref{psi_n}) and (\ref{phi_n}). It reads
as
\begin{eqnarray}
\tilde{\omega}_n=\frac{1}{2}\left[\,\Omega_n^F\cos{2\varphi_n}+|C_n^F|\sin{2\varphi_n})\,\right]
=\frac{1}{2}\sqrt{(\Omega_n^F)^2+|C_n^F|^2}\,\,\,.
\end{eqnarray}
By struggling with Bessel functions we get the remarkably nice
result (The details have been included in Appendix B)
\begin{eqnarray}
\tilde{\omega}_n(\tau)=\omega_n(\tau)\equiv\sqrt{n^2+\tilde{f}^2e^{-2\tau}}\,\,\,.\label{cancel}
\end{eqnarray}
Then the fermionic part of the Hamiltonian can be diagonalized as
\begin{eqnarray}\label{HF}
H_F=H_{F0}(\tau)+\frac{1}{\alpha'p^+}\sum_{n=1}^\infty\,
    \omega_n(\tau)\left[B_n^\dag(\tau)B_n(\tau)
    +\tilde{B}_n^\dag(\tau)\tilde{B}_n(\tau)-1\right].
\end{eqnarray}
Just like the bosonic case, it also looks like the Hamiltonian of a
free 2-d field theory with time-dependent mass.

Taking account of the bosonic sector, the full light-cone Hamiltonian takes the form
\begin{eqnarray}\label{H}
H&=&H_0(\tau)+\frac{1}{\alpha'p^+}\sum_{n=1}^\infty\,
    \omega_n(\tau)\left[\,\,A_n^\dag(\tau)A_n(\tau)
    +\tilde{A}_n^\dag(\tau)\tilde{A}_n(\tau)\right.\nonumber\\
&&\left.\hspace{5cm}+B_n^\dag(\tau)B_n(\tau)
    +\tilde{B}_n^\dag(\tau)\tilde{B}_n(\tau)\,\,\right]\,\,,
\end{eqnarray}
with
\begin{eqnarray}
H_0(\tau)=H_{B0}(\tau)+H_{F0}(\tau)\,\,.
\end{eqnarray}
We can see that the zero-point energy exactly cancels between the
bosons and the fermions. Just like the bosonic case, one may
define the fermionic spectrum formally, which is also
time-dependent. From the total Hamiltonian, it is obvious that the
spectrum is symmetric between the bosonic and fermionic
excitations.

In this and the last section, we have discussed the quantization
of the free bosonic and fermionic strings in the plane-wave
background with a linear null dilaton. In the lightcone gauge, the
calculation of  perturbative string amplitudes is simple formally.
In the case with constant $\lambda$, the spectrum is exactly the
same as the ones in maximally supersymmetric plane-wave case, and
the perturbative amplitudes in the lightcone gauge is also very
similar except the overall dependence on the string coupling. The
dependence on the string coupling shows that the perturbative
calculation is ill-defined in the strong coupling region. In the
case with constant RR field strength, it is more subtle.  When
$\tau\rightarrow -\infty$, not only the string coupling is very
strong, but also the energy of the excitations is huge. We wish
the perturbative amplitude calculation is still well-defined in
the weak coupling region.

\section{Quantum string mode creation}

Generically, in a time-dependent background, one may expect the
particle or string creation occurs from our knowledge of the quantum
field theory in curved spacetime. However, in a plane-fronted
background, due to the existence of null Killing vector, this would
not happen\cite{Gibbons}. Nevertheless, as pointed out in
\cite{90Horow}, there does exist the string mode creation. This
could be easily seen from the time-dependent Hamiltonian we have
obtained above. From the point of view of  two-dimensional quantum
field theory, we have a time-dependent potential, which induces the
transition between different modes of the string. See also
\cite{91Brooks,92Vega, 93Vega, 93Jofre}.

Now let us study quantum string mode creation in our background
(\ref{BG}).  In general, given a pp-wave background with
asymptotically flat region at $\tau=+\infty$, a string starting in a
certain state at $\tau=+\infty$ evolves back with time and maybe end
up in a different sate. Equivalently, one may reverse the
orientation of time (which is equal to change the sigh of $c$ in
$\phi=-cx^+$) and interpret this as an evolution from some excited
state to the vacuum at $\tau=+\infty$. The main reason is that the
string may interact with the pp-wave background to have extra
internal excitations.

In the following we will consider whether an observer in the ``in"
vacuum $|0\rangle_\infty$ at $\tau=+\infty$ will see string mode
creation. Here the vacuum $|0\rangle_\infty$ is the Fock space state
which is annihilated by the operators $\alpha^I_n$ ,
$\tilde{\alpha}^I_n$ defined in (\ref{OC}) and $\beta^I_n$ ,
$\tilde{\beta}^I_n$ defined in (\ref{FAC}). We shall start with the
string in $|0\rangle_\infty$ state at $\tau=+\infty$ and study how
this state should evolve back to $\tau=-\infty$. In other words, we
will calculate the probability of a string state $|n, \tau=-\infty>$
transiting into the vacuum state at $\tau=\infty$.

First let us see the expectation value of the ``oscillator number''
operator that appears in the bosonic Hamiltonian (\ref{[A]})
\begin{eqnarray}
\bar{N}^B_n(\tau)&\equiv&\,_\infty\langle0|[\,A_n^{I\dag}(\tau)A^I_n(\tau)
         +\tilde{A}_n^{I\dag}(\tau)\tilde{A}^I_n(\tau)]|0\rangle_\infty\,\nonumber\\
                 &=&2\,d\,g^{\ast}_n(\tau)g_n(\tau)\,\,\,,
\end{eqnarray}
where we have used (\ref{DA}) and $d=8$, the range of index $I$.
Inserting the definition (\ref{fg}) of $g_n(\tau)$, we find
\begin{eqnarray}\label{NB}
\bar{N}^B_n(\tau)=d\left[\frac{\Omega^B_n}{2\,\omega_n}-1\right]\,\,.
\end{eqnarray}
Using the same method, we get  the expectation value of the
``oscillator number'' operator in the fermionic Hamiltonian
(\ref{HF})
\begin{eqnarray}
\bar{N}^F_n(\tau)&\equiv&\,_\infty\langle0|[\,B_n^{I\dag}(\tau)B^I_n(\tau)
         +\tilde{B}_n^{I\dag}(\tau)\tilde{B}^I_n(\tau)]|0\rangle_\infty\,\nonumber\\
                 &=&d\left[1-\frac{\Omega^F_n}{2\,\omega_n}\right]\,\,\,,
\end{eqnarray}
Therefore the total number of created oscillator modes is
\begin{eqnarray}\label{TotalN}
\bar{N}_T(\tau)&=&\sum^{\infty}_{n=1}\left[\bar{N}^B_n+\bar{N}^F_n\right]\,\,,\nonumber\\
               &=&d\sum^{\infty}_{n=1}\left[\frac{\Omega^B_n-\Omega^F_n}{2\,\omega_n}\right]\,\,.
\end{eqnarray}
As $\tau\rightarrow+\infty$,
\begin{eqnarray}
\omega_n(\tau)\sim n\quad\,,\quad\Omega^B_n(\tau)\sim
2\,n\quad\,,\quad\Omega^F_n(\tau)\sim 2\,n\,\,,
\end{eqnarray}
we can see
\begin{eqnarray}
\bar{N}_T(\tau)\sim 0\,\,.
\end{eqnarray}
This is what we want since now $A^I_n(\tau)$, $\tilde{A}_n(\tau)$
are the same as $\alpha^I_n$, $\tilde{\alpha}^I_{n}$ and
$B_n(\tau)$, $\tilde{B}_n(\tau)$ are the same as $\beta_n$,
$\tilde{\beta}_{n}$; we do not expect to see string modes creation
in flat space. As $\tau\rightarrow-\infty$,
\begin{eqnarray}
\bar{N}^B_n(\tau)\sim\frac{2\,d\,e^{-n\pi}}{e^{n\pi}-e^{-n\pi}}\quad\,,\quad
\bar{N}^F_n(\tau)\sim\frac{2\,d\,e^{-n\pi}}{e^{n\pi}+e^{-n\pi}}\,\,,
\end{eqnarray}
the total number of created oscillator modes is
\begin{eqnarray}
\bar{N}_T(\tau)&=&4d\sum^{\infty}_{n=1}\frac{1}{e^{2n\pi}-e^{-2n\pi}}\,\nonumber\\[0.2cm]
               &\sim&0.06\,\,.
\end{eqnarray}
 Therefore there is nearly no string mode creation as
$\tau\rightarrow-\infty$. 

Here is an effective way to understand the above result. As pointed
out in \cite{90Horow}, the problem could be restated as a quantum
mechanical problem. Let us focus on the bosonic sector for brevity.
>From the equation of motion, we know that the $T_n(\tau)$ in
(\ref{Tn}) should satisfy the equation
 \be
 \p^2_\tau\,T_n+(\,n^2+\tilde{f}^2e^{-2\tau})\,T_n=0.
 \ee
Replacing $\tau$ by $x$ and $T_n$ by $\psi$, the above equation
takes a form of one-dimensional Schrodinger equation for a particle
with energy $n^2$ in  a potential $-\tilde{f}^2e^{-2\tau}$. The
problem of calculating  the number of the creating modes reduces to
the problem of calculating the reflective amplitude in this
one-dimensional system. Here the in-coming wave is a plane-wave from
the $x=\infty$ and there is a very deep well near the $x=-\infty$.
Obviously the probability of the reflection is very low. This is
consistent with the above calculation. Note that the large tidal
force near the $\tau=-\infty$ is attractive rather than repulsive,
this makes the difference. In the repulsive case, one has a large
potential barrier in the corresponding quantum mechanical system,
and the reflective amplitude is large so that the string mode
transition is violent.

Furthermore, it is remarkable that the linear null dilaton does
not affect the discussion on the string mode creation. As shown in
\cite{90Horow}, the effect of the dilaton in the equation of the
string evolution is in a form of double derivatives. For the
linear dilaton, it has no effect on the string evolution.

\section{Concluding remarks}

In this paper, we studied the perturbative string theory in a
time-dependent plane-wave background with a constant RR field
strength and a linear null dilaton. In the light-cone gauge, we
obtain  two-dimensional free massive field theories with
time-dependent masses in both the bosonic and fermionic sectors.
Following the standard canonical quantization, we obtained a
time-dependent Hamiltonian with vanishing zero-point energy. The
spectrum also is time-dependent and symmetric between bosonic and
fermionic excitations. Finally we investigated the string mode
creation in our background and found that it is negligible.

It is remarkable that the symmetric spectrum of the bosonic and
fermionic excitations does not come from the spacetime
supersymmetries. As it is well-known, even in the case with
supernumerary supersymmetries, the excitation spectrum of the
bosonic sector and fermionic sector is in general different. This
is because that  the masses of the transverse bosons in the action
depends on the form of the metric and could be different from each
other. On the other hand, the fermionic action depends only on the
RR field strength in the lightcone gauge so that the masses of the
fermionic excitations are the same among the transverse
directions. Therefore the symmetric spectrum found in this paper
is from the special choice of the metric in (\ref{BG}). One may
consider the following background
 \begin{eqnarray}
&&ds^2 = -2dx^+dx^--\sum_I\lambda_I(x^+)\,x_I^2\,dx^+dx^+ +dx^Idx^I\,,\nonumber \\
&&\phi=\phi(x^+)\,,\quad\quad(F_5)_{+1234}=(F_5)_{+5678}=2f.
\end{eqnarray}
Here $\lambda_I$ could be different from each other. The conformal
invariance condition requires that
 \begin{eqnarray}
\sum_I\lambda_I=-2\phi''+8f^2e^{2\phi}.
\end{eqnarray}
It is not hard to find that such backgrounds still keep sixteen
supersymmetries characterized by the Killing spinor $\epsilon$
satisfying $\Gamma^+\epsilon=0$.  The string theory in these
backgrounds could be solved exactly in the lightcone gauge, at
least for the cases with a constant RR field strength or the
constant $\lambda_I$. The straightforward calculation shows that
the transverse bosons in the bosonic action, whose form is similar
to (\ref{BA}), have different masses proportional to $\lambda_I$.
Therefore, the bosonic excitations are different from each other.
It is impossible to have symmetric spectrum between bosonic and
fermionic sector any more. As a consequence, the zero-point energy
cannot be cancelled exactly.

Our study is just the first step to understand string theory in
such backgrounds. As we have seen, there exists the cosmological
singularity at $x^+=-\infty$, where the string coupling is
divergent. This indicates that the perturbative string description
breaks down and we need other degrees of freedom to describe the
physics there. In the strong coupling region, the nonperturbative
degrees of freedom are essential to describe the physics. In the
spirit of BFSS conjecture of M-theory\cite{BFSS}, the matrix
model, which describe the dynamics of the partons, is a natural
candidate. There are a few recent papers studying the matrix
models in various time-dependent backgrounds\cite{Verlinde, Miao,
Bin, Das}. In particular, in \cite{Bin,Das}, the 1/2-BPS
plane-wave backgrounds of the 11D supergravity have been found and
the corresponding matrix models have been constructed. It would be
very interesting to construct the matrix description of IIB
backgrounds in this paper. This is under our investigation.
Comparing with the cases in \cite{Bin,Das}, one advantage of the
backgrounds discussed here is that they are solvable. With the
perturbative study in this paper, it would be illuminating to see
how the matrix degrees of freedom at the big bang are frozen to
those of the perturbative strings in the late time.

\section*{Acknowledgements}

We are grateful to Miao Li for reading the draft and giving
valuable suggestions. BC would like to thank ICTS, where part of
the work was done during his visit. This work was supported by
CNSF Grants and the key Grant project of Chinese Ministry of
Education (No. 305001).

\appendix
\section{Notation}
In this appendix we collect our convention of the 10d Gamma matrices
with respect to the vierbeins, not to the coordinates. Therefore the
Gamma matrices defined here all carry hatted indices
$\hat{\mu},\hat{\nu}$ etc. They satisfy the standard anticommutators
\begin{eqnarray}
\{\Gamma^{\hat{\mu}},\Gamma^{\hat{\nu}}\}=2\,\eta^{\hat{\mu}\hat{\nu}},\quad
\hat{\mu},\hat{\nu}=+,-,I,\quad I=1,2,\ldots,8.
\end{eqnarray}
where
\begin{eqnarray}
\eta^{\hat{\mu}\hat{\nu}}=\begin{pmatrix}0&-1&0\\-1&0&0\\0&0&\,\,\mathbf{1}_8\end{pmatrix},
\end{eqnarray}
i.e. $(\Gamma^{\hat{+}})^2=(\Gamma^{\hat{-}})^2=0,\,\{\Gamma^{\hat{+}},\Gamma^{\hat{-}}\}=-2,\,
\{\Gamma^{\hat{I}},\Gamma^{\hat{J}}\}=2\delta^{IJ}$.
We also define $\Gamma^0$ and $\Gamma^9$ by $\Gamma^{\hat{\pm}}=(\Gamma^0\pm\Gamma^9)/\sqrt{2}$.
We can choose the representation of Gamma matrices as
\begin{eqnarray}
\Gamma^{\hat{+}}=i\begin{pmatrix}0&\sqrt{2}\\0&0\end{pmatrix}\,,\quad
\Gamma^{\hat{-}}=i\begin{pmatrix}0&0\\\sqrt{2}&0\end{pmatrix}\,,\quad
\Gamma^{\hat{I}}=\begin{pmatrix}\gamma^I&0\\0&-\gamma^I\end{pmatrix}\,.
\end{eqnarray}
The 10d chiral matrix $\Gamma$ is
\begin{eqnarray}
\Gamma=\begin{pmatrix}\gamma&0\\0&-\gamma\end{pmatrix}\,.
\end{eqnarray}
Here $\gamma^I$'s satisfy $\{\gamma^I,\gamma^J\}=2\delta^{IJ}$,
which is a representation of $SO(8)$ Clifford algebra. We can choose
them as real and symmetric $8\times8$ matrices. $\gamma$ is the
chiral matrix of $\gamma^I$'s. In our representations we have
$C=\Gamma^0$ with $C$ being the 10d charge conjugation matrix.
Therefore the components of Majorana spinors in this representation
are all real. This property renders many things simple. We also define
$\Gamma_{\hat{\mu}}=\eta_{\hat{\mu}\hat{\nu}}\Gamma^{\hat{\nu}}$, so
\begin{eqnarray}
\Gamma_{\hat{+}}=-\,\Gamma^{\hat{-}},\quad \Gamma_{\hat{-}}=-\,\Gamma^{\hat{+}},\quad \Gamma_{\hat{I}}=\Gamma^{\hat{I}}.
\end{eqnarray}

The fermionic light-cone gauge condition $\Gamma^{\hat{+}}\theta^A=0$ implies, in this representation,
that the latter 16 components of the fermionic coordinates $\theta^A$ all vanish, i.e.
\begin{eqnarray}
\theta^A=\begin{pmatrix}\vartheta^A\\0\end{pmatrix}\,,\quad A=1,2\,.
\end{eqnarray}
Then the positive 10d chirality $\Gamma\theta^A=\theta^A$ implies that
$\gamma\,\vartheta^A=\vartheta^A$.
Since we can choose the representation of $\gamma^I$ properly such that $\gamma=\mathrm{diag}(1,-1)$,
so $\gamma\,\vartheta^A=\vartheta^A$ is followed by the consequence that only the first 8 components
of $\vartheta^A$, also $\theta^A$, are nonzero. Therefore we end up with two $SO(8)$ Majorana-Weyl
spinors both in the same chiral representation.
For more aspects of this representation, see the reference \cite{03Sadri}.

\section{Proof of two propositions}
In this section we give the proof of two claims
\begin{eqnarray}
&&(1)\quad
  \Omega^{F}_n=-2\,i\left[W_n\partial_{\tau}W^{\ast}_n
  +\tilde{W}_n\partial_{\tau}\tilde{W}^{\ast}_n\right]\,\,\,\mathrm{is\,\, positive\,\,definitely.}\nonumber\\[0.2cm]
&&(2)\quad
  (\Omega^{F}_n)^2+|C_n^F|^2=4n^2+4u^2\,,\quad u=\tilde{f}e^{-\tau}.\nonumber
\end{eqnarray}
For convenience we first list some properties of the Bessel function $J_\nu(z)$.
The Bessel function $J_\nu(z)$ is the solution of following the Bessel equation (about $y(z)$)
\begin{eqnarray}
\frac{1}{z}\,\frac{d}{dz}\left(z\,\frac{dy}{dz}\right)+\left(1-\frac{\nu^2}{z^2}\right)y=0\,.\label{DiffEq}
\end{eqnarray}
The Bessel function $J_\nu(z)$ has the following relations
\begin{eqnarray}
zJ_{\nu-1}&=&\nu J_\nu+zJ'_\nu\,,\label{D1+}\\
zJ_{\nu+1}&=&\nu J_\nu-zJ'_\nu\,.\label{D1-}
\end{eqnarray}
>From these two formulae it has
\begin{eqnarray}
\frac{2\nu}{z}J_\nu&=&J_{\nu-1}+J_{\nu+1}\,,\label{D2+}\\
2\,J'_\nu&=&J_{\nu-1}+J_{\nu+1}\,.\label{D2-}
\end{eqnarray}
The Wronskian determinant of $J_{\pm\nu}(z)$ is
\begin{eqnarray}
W[J_{\nu},J_{-\nu}]\equiv J_\nu J'_{-\nu}-J'_\nu J_{-\nu}=-\frac{2\sin{\nu\pi}}{\pi z}\,.\label{Wroskian}
\end{eqnarray}
By using of (\ref{D1+}) and (\ref{Wroskian}) we have
\begin{eqnarray}
J_{-\frac{1}{2}+in}(z)\,J_{-\frac{1}{2}-in}(z)+J_{\frac{1}{2}+in}(z)\,J_{\frac{1}{2}-in}(z)
=\frac{2\cosh{n\pi}}{\pi z}\,\,.\label{1/2}
\end{eqnarray}
In general the parameter $\nu$ and the argument $z$ are both complex numbers.
For real argument $z=u\in\mathbb{R}$ we have $J^{\,^\ast}_\nu(u)=J_{\nu^\ast}(u)$.

\subsection{Proof of (1)}
Firstly we can simplify the expression of $\Omega^{F}_n$ as
\begin{eqnarray}
\Omega^{F}_n&=&\frac{2i\pi u}{\cosh{n\pi}}\left\{\frac{\cosh{n\pi}}{2\pi u}+G(u)\right\}\nonumber\\[0.2cm]
&\equiv&\frac{2i\pi u}{\cosh{n\pi}}\left\{\frac{\cosh{n\pi}}{2\pi u}
  +\frac{u}{2}\left[J_{-\frac{1}{2}+in}(u)\,J'_{-\frac{1}{2}-in}(u)
  +J_{\frac{1}{2}+in}(u)\,J'_{\frac{1}{2}-in}(u)\right]\right\}.\label{Omega-1}
\end{eqnarray}
Here we have used (\ref{1/2}) and defined the function $G$ as in the second line.
The real part of $G$ is
\begin{eqnarray}
\mathrm{Re}\,G&=&\frac{1}{2}\,\left\{G+G^\ast\right\}\nonumber\\
&=&\frac{u}{4}\,\frac{d}{du}\left\{|J_{-\frac{1}{2}+in}|^2+|J_{\frac{1}{2}+in}|^2\right\}\nonumber\\[0.1cm]
&=&-\frac{\cosh{n\pi}}{2\pi u}\,,
\end{eqnarray}
which cancels the first term in the first line of (\ref{Omega-1}).
Then we have proved that $\Omega^{F}_n$ is real. Secondly we should
show that the remaining part
\begin{eqnarray}
\Omega^{F}_n=-\frac{2\,\pi\,u}{\,\cosh{n\pi}\,}\,\,\mathrm{Im}\,G
\end{eqnarray}
is positive definitely for all $u>0$. To do so we consider of the derivative of $\mathrm{Im}G$.
It is not difficult to get that
\begin{eqnarray}
\frac{d}{du}\,\mathrm{Im}G=\frac{n}{2u}\left\{|J_{-\frac{1}{2}+in}|^2+|J_{\frac{1}{2}+in}|^2\right\}>0\,,\quad
\mathrm{for}\,\, u>0\,.
\end{eqnarray}
Here we have used the Bessel equation (\ref{DiffEq}) to eliminate the second derivative terms.
The limit of $\mathrm{Im}G$ as $u\rightarrow+\infty$ is $\frac{1-e^2}{2\pi e}<0$, which can be obtained
by the standard formulae of asymptotic behaviors of Bessel functions. Then $\mathrm{Im}G$ is negative definitely.
Therefore we have prove that $\Omega^{F}_n$, although is complex formally, is not only real but also positive
definitely for all $u>0$.

\subsection{Proof of (2)}
Here we give the details of the proof of (\ref{cancel}) which states that
\begin{eqnarray}
(\Omega_n^F)^2+|C_n^F|^2=4n^2+4u^2\,,\quad u=\tilde{f}e^{-\tau}>0\,.\label{cancelApp}
\end{eqnarray}
Using the definitions (\ref{Omega^F}) and (\ref{C^F}) of $\Omega_n^F$ and $C_n^F$,
it is not difficult to know that
\begin{eqnarray}
(\Omega_n^F)^2+|C_n^F|^2=\frac{4\pi{u}^2}{\,\cosh{n\pi}\,}(A+B+C)
\end{eqnarray}
with
\begin{eqnarray}
A&=&\frac{1}{8u}\left\{\,\left|J_{-\frac{1}{2}+in}(u)\right|^2+\left|J_{\frac{1}{2}+in}(u)\right|^2\right\}
    =\,\frac{\,\cosh{n\pi}\,}{4\pi{u}^2}\,,\\[0.1cm]
B&=&\frac{1}{4}\,\frac{d}{du}
    \left\{\,\left|J_{-\frac{1}{2}+in}(u)\right|^2+\left|J_{\frac{1}{2}+in}(u)\right|^2\right\}
    =\,-\frac{\,\cosh{n\pi}\,}{2\pi{u}^2}\,,\\[0.1cm]
C&=&\frac{u}{2}\left\{\,\left|J'_{-\frac{1}{2}+in}(u)\right|^2+\left|J'_{\frac{1}{2}+in}(u)\right|^2\right\}\,.
\end{eqnarray}
For the second equalities of function $A$ and $B$ we have used the identity (\ref{1/2}).
To simplify the function $C$ we utilize the formula (\ref{D1+}) and (\ref{D1-}) that
\begin{eqnarray}
J'_{-\frac{1}{2}\pm in}&=&\frac{\,\,-\frac{1}{2}\pm in}{u}\,J_{-\frac{1}{2}\pm in}-J_{\frac{1}{2}\pm in}\,,\\[0.2cm]
J'_{\,\,\frac{1}{2}\pm in}&=&-\,\frac{\,\frac{1}{2}\pm in}{u}\,J_{\frac{1}{2}\pm in}+J_{-\frac{1}{2}\pm in}\,,
\end{eqnarray}
and the formula (\ref{1/2}), then we have
\begin{eqnarray}
C&=&\frac{\,\,1+4n^2+4u^2\,}{4\pi u^2}\,\,.
\end{eqnarray}
Put all together we have proved what we want, i.e.
\begin{eqnarray}
(\Omega_n^F)^2+|C_n^F|^2=4n^2+4u^2\,,\quad u=\tilde{f}e^{-\tau}>0\,.
\end{eqnarray}
Therefore the zero-point energy really cancels between bosons and fermions.


\begin{thebibliography}{99}

\bibitem{GLS}H. Liu, G. Moore and N. Seiberg, ``Strings in a
         time-dependent orbifold", JHEP 0206 (2002) 045 [hep-th/0204168].
\bibitem{Cornalba}L. Cornalba and M. S. Costa, ``A New Cosmological Scenario
         in String Theory", Phy. Rev. D66 (2002) 066001 [hep-th/0203031].
         ``Time dependent orbifolds and string cosmology",
         Fortsch. Phys. 52, 145 (2004) [hep-th/0310099].
\bibitem{HP}G. T. Horowitz and J. Polchinski, ``Instability of Spacelike and Null Orbifold
         Singularities", Phys.Rev. D66 (2002) 103512 [hep-th/0206228].

\bibitem{Lawrence}A. Lawrence, ``On the instability of 3d null
        singularities", JHEP 0211 (2002) 019 [hep-th/0205288].

\bibitem{GLS1}H. Liu, G. Moore and N. Seiberg, ``Strings in
         time-dependent orbifolds", JHEP 0210 (2002) 031 [hep-th/0206182].
\bibitem{Kutasov}M. Berkooz, B. Craps, D. Kutasov, and G. Rajesh,
         ``Comments on Cosmological Singularities in String Theory", JHEP
         0303 (2003) 031 [hep-th/0212215].
\bibitem{Berkooz}M. Berkooz, Z. Komargodski, D. Reichmann and V.
         Shpitalnik, ``Flow of Geometries and Instantons on the Null
         Orbifold" [hep-th.0507067].
\bibitem{Sen}A. Sen, ``Rolling Tachyon", JHEP 0204 (2002) 048 [hep-th/0203211].
          ``Open-closed duality at tree level", Phys. Rev.
         Lett. 91 (2003) 181601 [hep-th/0306137].
\bibitem{Silverstein}J. McGreevy and E. Silverstein, ``The Tachyon at the End of the
         Universe", [hep-th/0506130].
\bibitem{Verlinde}B. Craps, S. Sethi and E. Verlinde,  ``A Matrix Big Bang", [hep-th/0506180].
\bibitem{Miao}Miao Li, ``A class of cosmological matrix models", [hep-th/0506260]. \\
        Miao Li and Wei Song, ``Shock waves and cosmological matrix
        models", [hep-th/0507185].
\bibitem{Bin}Bin Chen, ``The Time-dependent Supersymmetric Configurations in M-theory and Matrix
        Models", [hep-th/0508191].
\bibitem{Das}S. R. Das and J. Michelson, ``pp Wave Big Bangs:
        Matrix Strings and Shrinking Fuzzy Spheres", [hep-th/0508068].
\bibitem{she}J.-H. She ``A Matrix Model for Misner Universe and Closed String Tachyons'', [hep-th/0509067].
\bibitem{o1}R.-G. Cai, J.-X. Lu and N. Ohta, ``NCOS and D-branes in Time-dependent
        Backgrounds'', Phys. Lett. B551 (2003) 178 [hep-th/0210206].
\bibitem{o2}N. Ohta, M. Sakaguchi, ``Uniqueness of M-theory PP-Wave Background with Extra
        Supersymmetries'', Phys.Rev. D69 (2004) 066006 [hep-th/0305215].
\bibitem{Liouville}N. Seiberg, ``Notes on Quantum Liouville Theory and Quantum Gravity",
Prog.Theor.Phys.Suppl.102:319-349,1990.
\bibitem{2Dstring}I.R. Klebanov, ``Strings in two-dimensions",
[hep-th/9110041].\\
P. Ginsparg, G. Moore, ``Lectures on 2D gravity and 2D string
theory", [hep-th/9304011].
\bibitem{Matrix}I. R. Klebanov, J. Maldacena, N. Seiberg,``D-brane Decay in Two-Dimensional String
Theory", JHEP 0307 (2003) 045, [hep-th/0305159].\\
J. McGreevy, J. Teschner and H. Verlinde, ``Classical and Quantum
D-branes in 2D String Theory", JHEP 0401 (2004) 039,
[hep-th/0305194].\\
J. McGreevy and H. Verlinde, ``Strings from Tachyons", JHEP 0312
(2003) 054, [hep-th/0304224].

\bibitem{90Horow}G. T. Horowitz and A. R. Steif, `` Strings In Strong Gravitational
                 Fields,'' Phys. Rev. D 42 (1990) 1950.`` Space-Time Singularities
                 In String Theory,'' Phys. Rev. Lett. 64 (1990) 260.

\bibitem{Metsaev01} R. R. Metsaev, ``Type IIB Green-Schwarz superstring in plane wave Ramond-Ramond background,''
                Nucl.Phys. B625 (2002) 70 [hep-th/0112044].
\bibitem{MT02}R. R. Metsaev and A. A. Tseytlin, ``Exactly solvable model of superstring in Ramond-Ramond
                plane wave background,'' Phys.Rev. D65 (2002) 126004 [hep-th/0202109].
\bibitem{Blau}M. Blau, J. Figueroa-O'Farrill, C. Hull and G.
        Papadopoulos, ``A new maximally supersymmetric background of IIB
        superstring theory ", JHEP 0201 (2002) 047 [hep-th/0110242].
        ``Penrose limits and maximal supersymmetry", Class.Quant.Grav. 19
        (2002) L87 [hep-th/0201081].
\bibitem{BMN}D. Berenstein, J. Maldacena and H. Nastase, ``Strings
        in flat space and pp-waves from $N=4$ super-Yang-Mills", JHEP 0204 (2002) 013 [hep-th/0202021].
\bibitem{02Tsey} G. Papadopoulos, J. G. Russo and A. A. Tseytlin,
               `` Solvable model of strings in a time-dependent palne-wave background,''
                Class.Quant.Grav. 20 (2003) 969 [hep-th/0211289].

\bibitem{Marolf}D. Marolf and L. A. Pando Zayas ``On the Singularity Structure and Stability of Plane Waves'',
        JHEP 0301 (2003) 076 [hep-th/0210309].
\bibitem{Cvetic1}M. Cvetic, H. Lu, C.N. Pope, `` M-theory PP-waves, Penrose Limits and Supernumerary
        Supersymmetries'', Nucl.Phys. B644 (2002) 65 [hep-th/0203229].
\bibitem{Cvetic2}M. Cvetic, H. Lu, C.N. Pope, K.S. Stelle, ``Linearly-realised Worldsheet Supersymmetry
        in pp-wave Background'', Nucl.Phys. B662 (2003) 89 [hep-th/0209193].

\bibitem{02Russo} J. G. Russo and A. A. Tseytlin, ``A class of exact pp-wave string models with interacting
                light-cone gauge actions,'' JHEP 0209 (2002) 035 [hep-th/0208114].
\bibitem{Gibbons}G. W. Gibbons, ``Quantized fields propagating in
        plane wave spacetime", Commun. Math. Phys. 45 (1975) 191.
\bibitem{Guven}R. Guven, Phys. Lett. B191 (1987)275.
\bibitem{Amati}D. Amati and C. Klimcik, Phys. Lett. B219 (1989)443.

\bibitem{91Brooks}R. Brooks, ``Plane Wave Gravitons, Curvature Singularities And String
                Physics,'' Mod. Phys. Lett. A6 (1991) 841.
\bibitem{92Vega}H. J. de Vega and N. Sanchez, ``Strings Falling Into Space-Time Singularities,'' Phys.
                Rev. D 45 (1992) 2783.
\bibitem{93Vega}H. J. de Vega, M. Ramon Medrano and N. Sanchez, ``Classical
                and quantum strings near space-time singularities: Gravitational plane waves with arbitrary
                polarization,'' Class. Quant. Grav. 10 (1993) 2007.
\bibitem{93Jofre}O. Jofre and C. Nunez, ``Strings In Plane Wave Backgrounds Revisited,'' Phys. Rev.
                D50 (1994) 5232 [hep-th/9311187].
\bibitem{03Sadri} D. Sadri and M. M. Sheikh-Jabbari, ``The Plane-Wave/Super Yang-Mills Duality,''
                Rev.Mod.Phys. 76 (2004) 853 [hep-th/0310119].

\bibitem{BFSS} T. Banks, W. Fischler, S.H. Shenker, L. Susskind,
``M Theory As A Matrix Model: A Conjecture", Phys.Rev. D55 (1997)
5112-5128, [hep-th/9610043].


\end{thebibliography}
\end{document}